\documentclass[graybox]{svmult}

\usepackage{amssymb}
\usepackage{amsmath}

\usepackage{url}

\usepackage{mathptmx}       
\usepackage{helvet}         
\usepackage{courier}        
\usepackage{type1cm}        
\usepackage{makeidx}         
\usepackage{graphicx}        
\usepackage{multicol}        
\usepackage[bottom]{footmisc}

\usepackage[authoryear]{natbib}
\usepackage{subfigure}

\makeindex             
                       
\def\<<{{\ll}}
\def\>>{{\gg}}

\def\spose#1{\hbox to 0pt{#1\hss}}
\def\ltwig{\mathrel{\spose{\lower 3pt\hbox{$\mathchar"218$}}
     \raise 2.0pt\hbox{$\mathchar"13C$}}}
\def\gtwig{\mathrel{\spose{\lower 3pt\hbox{$\mathchar"218$}}
     \raise 2.0pt\hbox{$\mathchar"13E$}}}
\def\+/-{{\pm}}
\def\=={{\equiv}}
\def\Rstar{R}
\def\Mstar{M}
\def\Lstar{L}

\def\Fvec{{\bf F}}
\def\vvec{{\bf v}}

\def\gvecrad{{\bf g}_{\rm rad}}
\def\rhat{{\bf {\hat r}}}
\def\ws{w_{\rm s}}

\def\kapedd{\kappa_{\rm Edd}}
\def\Teff{T_{\rm eff}}

\def\grad{g_{\rm rad}}

\def\Mdot{\dot M}

\def\solar{\odot}
\def\Msun{M_{\solar}}

\def\Rsun{R_{\solar}}
\def\Lsun{L_{\solar}}

\def\vesc{v_{\rm esc}}

\newcommand{\beq}{\begin{equation}}
\newcommand{\eeq}{\end{equation}}
\newcommand{\beqa}{\begin{eqnarray}}
\newcommand{\eeqa}{\end{eqnarray}}
\def\Qbar{{\overline Q}}

\def\nhvec{{\bf{\hat n}}}
\def\rhvec{{\bf{\hat r}}}

\newcommand{\Prad}{P_{\rm rad}}
\newcommand{\Pgas}{P_{\rm gas}}

\newcommand{\Bnu}{\ensuremath{B_{\nu}}}

\newcommand{\kapnu}{\ensuremath{\kappa_{\nu}}}
\newcommand{\kapF}{\ensuremath{{\bar \kappa}_{F}}}
\newcommand{\kapR}{\ensuremath{{\bar \kappa}_{R}}}
\newcommand{\kapP}{\ensuremath{{\bar \kappa}_{P}}}
\newcommand{\diff}{\ensuremath{\mathrm{d}}}

\begin{document}

\tableofcontents

\title*{Instabilities in the Envelopes and Winds of Very Massive Stars}
\author{Stanley P. Owocki}
\authorrunning{Instabilities in VMS Envelopes and Winds}
\institute{Stanley P. Owocki \at Department of Physics \& Astronomy, University of Delaware, Newark, DE 19716 USA, \email{owocki@udel.edu}
}
%
%
\maketitle

\abstract*{
The high luminosity of Very Massive Stars (VMS) means that radiative forces play an important, dynamical role both in  the structure and stability of their stellar envelope, and in driving strong stellar-wind mass loss. 
Focusing on the interplay of radiative flux and opacity, with emphasis on key distinctions between continuum vs.\ line opacity, this chapter reviews instabilities in the envelopes and winds of VMS.
Specifically, we discuss how:
1) the iron opacity bump can induce an extensive inflation of the stellar envelope;
2) the density dependence of mean opacity leads to strange mode instabilities in the outer envelope;
3) desaturation of line-opacity by acceleration of near-surface layers initiates and sustains a line-driven stellar wind outflow;
4) an associated line-deshadowing instability leads to extensive small-scale structure in the outer regions of such line-driven winds;
5) a star with super-Eddington luminosity  can develop extensive atmospheric structure from photon bubble instabilities, or from stagnation of flow that exceeds the ``photon tiring'' limit;
6) the associated porosity  leads to a reduction in opacity that can regulate the extreme mass loss of such continuum-driven winds.
Two overall themes are the potential links of such instabilities to Luminous Blue Variable (LBV) stars, and the potential role of radiation forces in establishing the upper mass limit of VMS.
}

\abstract{
The high luminosity of Very Massive Stars (VMS) means that radiative forces play an important, dynamical role both in  the structure and stability of their stellar envelope, and in driving strong stellar-wind mass loss. 
Focusing on the interplay of radiative flux and opacity, with emphasis on key distinctions between continuum vs.\ line opacity, this chapter reviews instabilities in the envelopes and winds of VMS.
Specifically, we discuss how:
1) the iron opacity bump can induce an extensive inflation of the stellar envelope;
2) the density dependence of mean opacity leads to strange mode instabilities in the outer envelope;
3) desaturation of line-opacity by acceleration of near-surface layers initiates and sustains a line-driven stellar wind outflow;
4) an associated line-deshadowing instability leads to extensive small-scale structure in the outer regions of such line-driven winds;
5) a star with super-Eddington luminosity  can develop extensive atmospheric structure from photon bubble instabilities, or from stagnation of flow that exceeds the ``photon tiring'' limit;
6) the associated porosity  leads to a reduction in opacity that can regulate the extreme mass loss of such continuum-driven winds.
Two overall themes are the potential links of such instabilities to Luminous Blue Variable (LBV) stars, and the potential role of radiation forces in establishing the upper mass limit of VMS.
}

\section{Background: VMS M-L Relation \& the Eddington Limit}
\label{sec:background}
\label{sec1}
A hallmark of very massive stars (VMS) is that they are very, very luminous.
For example, a star of a hundred solar masses typically has a luminosity that is of order a {\em million} times the solar luminosity.
This means that, from the realm of solar to very massive stars, the luminosity scales  roughly with the {\em cube} of the stellar mass, $\Lstar \sim M^3$ (justifying perhaps adding even a third ``very'' to ``luminous").
This is {\it not} (as sometimes inferred) a consequence of the core nuclear burning source of the stellar luminosity, but instead, as worked out by \citet{Eddington26} and others even before nuclear burning was fully understood, follows from the basic equations of stellar structure, namely the dual requirements of {\em hydrostatic} pressure support against stellar gravity, and {\em radiative transport} of energy from the interior to the surface (see \S \ref{sec3}).
As was also recognized (most notably by Eddington) from these early studies of stellar structure, this cubic scaling of luminosity with mass can not be maintained to arbitrarily large masses, essentially because at high luminosity the associated {\em radiation pressure} becomes significant in the star's gravitational support.

Radiation pressure is a consequence of the fact that, in addition to their important general role as carriers of energy, photons also have an associated momentum, set by their energy divided by the speed of light $c$.
The trapping of radiative energy within a star thus inevitably involves a trapping of its associated momentum, leading to an outward radiative force, or for a given mass, an outward {\em radiative acceleration} $g_{rad}$, that can compete with the star's gravitational acceleration $g$. 
For a local radiative energy flux $F$ (energy/time/area), the associated momentum flux (force/area, or pressure) is just $F/c$.
The material acceleration resulting from absorbing this radiation depends on the effective cross sectional area $\sigma$ for absorption, divided by the associated material mass $m$,
\beq
g_{rad} = \frac{\sigma}{m} \frac{F}{c} \equiv \frac{\kappa F}{c}
\, .
\label{eq:grad}
\eeq
The latter equality defines the {\em opacity} $\kappa = \sigma/m$, which is just a measure of the total effective absorption cross section per unit mass of absorbing material.

For a star of luminosity $\Lstar$, the radiative flux at some radial distance $r$ is just $F=\Lstar/4\pi r^2$. This gives the radiative acceleration  the same inverse-square radial decline as the stellar gravity, $g = GM/r^2$ (with $G$ the gravitation constant);
their ratio, generally referred to as the ``Eddington parameter'',  thus tends to be relatively constant, set by the ratio of luminosity to mass,
\beq
\Gamma \equiv \frac{g_{rad}}{g}
= \frac{\kappa F}{gc} 
= \frac{\kappa \Lstar}{4 \pi G \Mstar c} 
=   \Gamma_e  \frac{\kappa}{\kappa_e}
\approx  2.6 \times 10^{-5} \frac{\kappa}{\kappa_e} \frac{\Lstar/\Mstar}{\Lsun/\Msun} 
\, .
\label{eq:gamdef}
\eeq
The last two equalities provide scalings in terms  of the classical Eddington parameter, $\Gamma_e$, defined for the electron scattering opacity, $\kappa_e \equiv \sigma_{Th}/\mu_e$, where $\sigma_{Th} = 6.7 \times 10^{-25}$\,cm$^2$\,g$^{-1}$ is the Thompson cross section for free electron scattering, and $\mu_e $ is the mean mass per free electron. The latter scales with the Hydrogen mass $m_H$ rather than the much smaller electron mass, because even for free electrons, maintaining overall charge neutrality requires an effective coupling between electrons and the ions that are the main contributors to the material mass. For a fully ionized gas with Hydrogen mass fraction $X$, $\mu_e = 2 m_H/(1+X)$, giving $\kappa_e=0.2 (1+X) \approx 0.34$\,cm$^2$\,g$^{-1}$ for standard (solar) mass fraction $X \approx 0.7$.
Applying this in (\ref{eq:gamdef}), the last equality shows that, for a star with the solar luminosity to mass ratio $\Lsun/\Msun$, the electron Eddington parameter $\Gamma_{e \odot}$ is very small, implying that for such solar-type stars the electron scattering acceleration $g_e = \Gamma_e g$ is entirely negligible compared to gravity.

But if one assumes an overall cubic scaling of luminosity with mass, then 
\beq
\Gamma_e = \Gamma_{e \odot} (\Mstar/\Msun)^2
\, ,
\eeq
which would reach the classical Eddington limit $\Gamma_e = 1$ for a mass $M_{Edd} = \Msun /\sqrt{\Gamma_{e \odot}}$ $\approx 195 \Msun$.
Rather remarkably, this 
agrees quite well with modern empirical estimates for the most massive observed stars, which are in the range 150-300\,$\Msun$ \citep{Figer05, Oey05, Crowther10, Crowther12}. (See Chapters 2, 3 and 8.)

\begin{figure}[t]
\begin{center}
\includegraphics[scale=0.7]{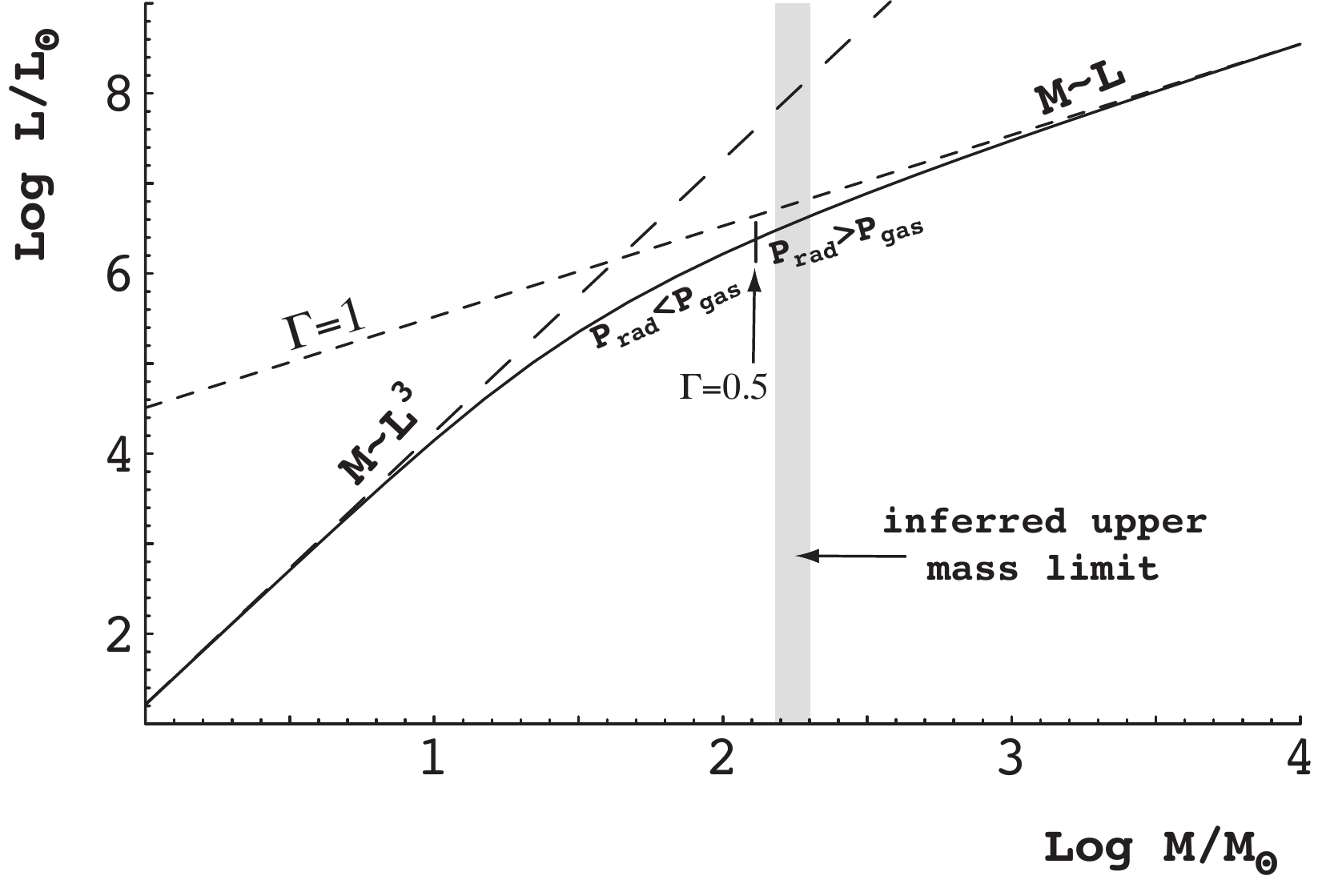}
\caption{
Log-log plot of the stellar luminosity $\Lstar$ vs.\
mass $\Mstar$ for the Eddington Standard Model scaling of equation (\ref{eq:esmgm}).
}
\label{fig:mlum}
\end{center}
\end{figure}

More complete analyses that account for the effect of radiation pressure in the hydrostatic support against gravity show that associated adjustments in the stellar structure formally allow gravitationally bound ($\Gamma < 1$) stars without {\em any} upper mass limit.
For example, for a homogenous star with solar Hydrogen mass fraction $X \approx 0.7$ and radially constant $\Gamma$,  the so-called ``Eddington Standard Model" (ESM) gives the scaling \citep{Eddington26},
\beq
\frac{\Gamma}{(1-\Gamma)^4} 
\approx \left ( \frac{\Mstar}{48 \Msun} \right )^2
\, .
\label{eq:esmgm}
\eeq
The $1-\Gamma$ factor comes from the reduction in effective gravity from radiation pressure; its presence as a quartic in the denominator represents a strong repeller against the Eddington limit, $\Gamma \rightarrow 1$. As illustrated in  figure \ref{fig:mlum}, this forces the low-mass cubic scaling $L \sim \Mstar^3$ to reduce to a linear scaling $L \sim \Mstar$ at large mass, always keeping below the limit.

The ESM assumption of a radially fixed $\Gamma$ also implies (see \S \ref{sec3.1}) a fixed ratio of the radiation pressure to gas pressure, 
\beq
\frac{\Prad}{\Pgas} \approx \frac{\Gamma}{1-\Gamma}
\, .
\label{eq:pgpresm}
\eeq
The case $\Gamma = 1/2$, with ESM mass $\Mstar = 136 \Msun$, has $\Prad = \Pgas$, and so marks the transition from gas to radiation as the dominant source of pressure support against gravity.
In analogy to having a heavier fluid be supported by a lighter one, an envelope with gravitational support that is predominately from radiation pressure is expected to be intrinsically unstable, sometimes dubbed the ``photon bubble instability"
\citep{Spiegel99, Shaviv98, Shaviv00, Shaviv01}.
If the nonlinear growth of this and related instabilities can sufficiently disrupt the stellar structure, for example inducing strong episodes of extensive mass loss,  it could be a key factor in setting an effective upper limit for stellar mass.

Indeed, the above-mentioned observationally inferred mass limit \citep{Crowther12} (Chapters 2, 3 and 8) is just above the ESM mass for transition 
to radiation pressure dominance.
For electron scattering opacity, the corresponding  ESM luminosity  
$\sim 5 \times 10^6 \Lsun$ ($M_{\rm bol} \sim -12$)
is near the luminosities of the intrinsically brightest observed stars, for example, $\eta~$Carinae or the `Pistol star', both of which show evidence for past episodes of strong mass loss.
These stars are prototypes of a ``giant eruption'' subclass of ``Luminous Blue Variable" (LBV) stars. 
On the Hertzsprung-Russell diagram they lie very near the Humphreys-Davidson's (H-D) limit \citep{Humphreys79}
that  delineates the most luminous observed stars.
A second subtype, the S-Doradus LBVs, can  occur a factor ten or more below the H-D limit;
in contrast to the strong brightenings of eruptive LBV's, they are characterized by year timescale variations in effective temperature, but with a  roughly {\em constant} bolometric magnitude.

Focusing on the interplay of opacity, radiative flux and gravity,  the remaining sections of this chapter review how the strong radiative acceleration in luminous VMS can lead to strong mass loss and induce instabilities in both their interior envelopes and stellar wind outflows. The goal is to provide a good physical basis for exploring the potential role of such radiative acceleration and the associated mass loss and instability for understanding both giant eruption and S-Doradus variability in LBVs, as well as for the inferred VMS upper mass limit.

Building on methods (\S \ref{sec2}), for estimating the flux-weighted mean opacity, \S {\ref{sec3} examines the effect of radiative forces on the structure and inflation of the hydrostatic, gravitationally bound stellar envelope.
We next (\S \ref{sec4}) write the general time-dependent equations for conservation of mass, momentum, and energy, and apply these (\S 4.2) to a linear perturbation analysis of `strange-mode' instabilities in the stellar envelope, and to write the basic equations for steady stellar wind outflow (\S 4.3).
Applying the latter to the standard case of a line-driven wind (\S 5), we derive steady solutions for the mass loss rate and wind velocity (\S 5.1), and then discuss  (\S\S 5.2-5.3) the extensive structure (clumping and porosity) that arises in time-dependent models that account for the strong {\em Line-Deshadowing Instability} (LDI) intrinsic to line-driving.
Finally, for the giant eruption LBVs with a super-Eddington luminosity, we review (\S 6) how the much stronger mass loss -- which can approach the ``photon tiring'' limit  for the luminosity to lift material out of the star's gravitational potential -- can be modeled in terms of a quasi-steady {\em continuum}-driven wind regulated by a porosity reduction in the effective opacity. 

\section{Mean Opacity Formulations}
\label{sec:meanop}
\label{sec2}

\subsection{Flux-Weighted Mean Opacity}
\label{sec:kapf}
\label{sec2.1}

In addition to the free electron scattering that provides a nearly fixed, frequency-independent (gray), baseline opacity, there are additional contributions associated with electron interactions with ions, namely through
free-free (f-f), bound-free (b-f) and bound-bound (b-b) processes.
The resonant nature of bound-bound transitions makes the associated {\em line} opacities very strong, and these, especially from complex heavy ions of iron-group elements, turn out to be particularly important in the structure of near-surface layers and in driving stellar wind mass loss.

To account for the strong frequency-dependence of the associated opacity, the simple expression (\ref{eq:grad}) for the radiative acceleration must now be generalized to the integral form,
\beq
g_{rad} (r)
= 2 \pi \int_{-1}^{1} d\mu \, \mu  \int_{0}^{\infty} \, d\nu \, \kappa_{\nu} \, 
I_{\nu} (\mu,r) / c
= \int_{0}^{\infty} \, d\nu \, \kappa_{\nu} \, F_{\nu}(r) / c
\, .
\label{eq:gradint}
\eeq
where the integrals are over frequency $\nu$ and radial direction cosine
$\mu = \nhvec \cdot \rhvec $ for radiation in vector direction $\nhvec$ with specific intensity $I_{\nu}$ 
and associated  opacity $\kapnu$.
If this opacity is {\em isotropic}\footnote{As discussed in \S \ref{sec5.1},  even when the opacity is formally isotropic in the atom's frame, a spherical wind expansion can lead to an anisotropy for {\em line} opacity in the stellar frame, through the directional dependence of the local velocity gradient.},
 the evaluation reduces to just the latter frequency integral  of $\kapnu$ times the associated energy flux $F_\nu$. 

To compute the radiative acceleration in terms of the local {\em bolometric} flux $F \equiv \int_0^\infty F_\nu d\nu $, as done in equation (\ref{eq:grad}), the appropriate opacity is now a {\em flux-weighted mean},
\beq
\kapF \equiv 
 \int_0^\infty  \frac{\kapnu \, F_\nu}{F} \, d\nu 
\, .
\label{eq:kapfdef}
\eeq
The dependence on local gas and radiation conditions means this opacity, and thus the Eddington parameter from equation (\ref{eq:gamdef}), both now generally vary with the local radius $r$.

While notationally convenient in connecting back to the simple gray opacity scalings, it is important to realize that computation of $\kapF$ can be very difficult, in principal requiring a {\em global} integral solution of the generally {\em nonlocal} radiative transport to obtain the frequency dependence of the local flux $F_\nu$, accounting for 
the frequency dependence of the opacity $\kappa_{\nu}$, as well as its dependence on ionization and excitation level of the absorbing ions.

\subsection{Planck Mean and its Dominance by Line Opacity}
\label{sec:planck}
\label{sec2.2}

To illustrate the potentially dominant importance of line opacity, let's first consider an {\em optically thin} limit in which absorbing ions are exposed fully to the local continuum radiation, unattenuated by any self-absorption within the lines. If we model this continuum flux spectrum as being given by the broad Planck blackbody function for the star's surface temperature, so that $F_\nu/F = B_\nu/B$, we see that the flux-mean opacity $\kapF$ of eqn.\ (\ref{eq:kapfdef}) is just given in terms of a {\em Planck mean} opacity, defined by
\beq
 \kapP \equiv \int_0^\infty 
  \frac{\kapnu \, B_\nu}{B} \, d\nu 
\, .
\label{eq:kappdef}
\eeq
As a direct mean,  $\kapP$ is dominated by the strongest opacity sources, namely 
by the cumulative contribution from individual spectral lines.
Relative to electron scattering, the line opacity of an individual line has the form,
\beq
\frac{\kappa_{\nu_i}}{\kappa_e} = \frac{n_i}{n_e} \, f_i \frac{\sigma_{cl}}{\sigma_e} \, \phi(\nu-\nu_i)
\, ,
\label{eq:kapibke}
\eeq
where $n_i$  and $n_e$ are the number densities of absorbing ions and electrons, and
the  line-profile function, $\phi (\nu - \nu_i)$, is narrowly peaked around the line-center frequency $\nu_i$, with unit normalization $\int_0^\infty \phi(\nu - \nu_i) \, d\nu = 1$.
The quantum mechanical oscillator strength,  $f_i$, corrects the frequency-integrated cross-section $\sigma_{cl}$ (with dimensions of area times frequency) obtained from the ``classical oscillator'' model of line absorption.
In terms of the classical electron radius $r_e \equiv e^2/m_e c^2$,  the  frequency-integrated line-cross-section is enhanced by the dimensionless factor
\beq
Q_{\lambda_i} 
\equiv \frac{\sigma_{cl}}{\nu_i \sigma_e} 
= \frac{\pi r_e c}{ \nu_i  \, 8 \pi r_e^2/3} 
= \frac{3}{8} \, \frac{\lambda_i} {r_e} 
= 1.5 \times 10^8 \, \frac{\lambda_i}{1000 \, \AA }
\, ,
\eeq
where $\lambda_i = c/\nu_i$ is the line-center wavelength, and the notation 
$Q_{\lambda_i}$ 
is chosen because it is related (by just a factor $\pi^2$) to the resonance {\em quality} $Q = \nu_i/\gamma_i$, with $\gamma_i$ the radiative damping rate
\citep{Gayley95}.
The very large numerical value for a sample UV wavelength stems from the resonance nature of line transitions, showing that, even when integrated over a broad frequency range that is much larger than the line width, a bound electron has an enormously larger cross section than a free electron.
The effect is somewhat analogous to blowing into a whistle vs.\ just open air; the response is very strong, but concentrated in a narrow frequency range near the resonance.

Upon integration over the individual line-profiles,
we can thus approximate the Planck opacity as a sum over the line index $i$, weighted by a factor $W_i = \nu_i B_{\nu_i}/B$ that reflects the blackbody strength at the line frequency $\nu_i$,
\beq
\frac{\kappa_{P}}{\kappa_e} = \sum_i \frac{n_i}{n_e} \, f_i Q_{\lambda_i}  W_i \equiv \Qbar \approx 2000 \left (\frac{Z}{0.02} \right )
\, ,
\label{eq:qbardef}
\eeq
where the notation and evaluation in the last two equalities are due to \citet{Gayley95}. For allowed transitions near the peak of the Planck function, both $f_i$ and $W_i$ are order unity; but for hot stars with high ionization and near-solar metallicity, only a relatively small fraction $ \sim 10^{-5}$ of electrons remain bound in metal ions, implying a similarly small cumulative abundance ratio $\sum_i n_i/n_e$ that counters the large resonance factor 
$Q_{\lambda_i}$,
leaving a more moderately strong average resonance quality $\Qbar \approx 2000$.

The upshot here is that, when unsaturated in this way, the radiative force from line-opacity can approach an upper limit that is substantially enhanced over that associated with electron scattering, with an associated Eddington parameter
\beq
\Gamma_{max} \approx \Qbar \Gamma_e
\, .
\label{eq:gammax}
\eeq
This means that for lines the requirement $\Gamma > 1$ to overcome gravity and drive a wind outflow can occur in any stars with electron Eddington parameters $\Gamma_e > 1/ \Qbar \approx 0.0005 $ (see \S \ref{sec5.1}).
For VMS with $\Gamma_e$ only a factor few below unity, it indicates the potential for strong {\em line-deshadowing instability} (LDI), with any optically thin portions of a wind outflow having radiative acceleration approaching a thousand times the acceleration of gravity (see \S\S \ref{sec5.2}-\ref{sec5.3}).

\subsection{ Rosseland  Opacity and Radiative Diffusion in Stellar Interior}
\label{sec:kapr}
\label{sec2.3}

Of course this full brunt of line opacity does not apply in the dense, opaque stellar interior because,  in diffusing outward, radiation preferentially leaks through the inter-line frequencies of lower continuum opacity, leaving only a significantly reduced flux within the lines.
Within such a {\em diffusion approximation} for radiation transport, the local frequency-dependent flux now scales as
\beq
F_\nu (r) \approx  -\left [ \frac{4 \pi}{3 \kapnu \rho} 
\frac{\partial \Bnu}{\partial T} \right ]\, 
\frac{dT}{dr}
\, ,
\label{eq:fnudiff}
\eeq
where  $\rho $ is the local density and $B_\nu (T)$ 
is the frequency-dependent Planck function for the interior temperature $T(r)$ at local radius $r$.
Application of (\ref{eq:fnudiff}) in (\ref{eq:kapfdef}) shows that in this diffusion limit, the flux-weighted opacity is now approximated by the so-called ``{\em Rosseland mean}'', 
\begin{equation}
\kappa_{R} \equiv
\frac{\int_{0}^{\infty} \frac{\partial \Bnu}{\partial T} \,\diff \nu}
     {\int_{0}^{\infty} \frac{1}{\kapnu} \frac{\partial \Bnu}{\partial T} \,\diff \nu}\
\, .
\label{eq:kaprdef}
\end{equation}
As a {\em harmonic mean},  $\kappa_R$ is  dominated by the weaker components of the opacity, with  generally little relative contribution from individual spectral  lines\footnote{An exception is when the spectral density of lines become high enough to lead to an effective ``line-blanketing'' effect, as occurs in the  iron opacity bump discussed in \S \ref{sec:opal}.}.
The numerator can be readily evaluated by taking the temperature derivative outside the frequency integral,
\beq
\int_{0}^{\infty} \frac{\partial \Bnu}{\partial T} \,\diff \nu
=
\frac{\partial }{\partial T} \int_{0}^{\infty} \Bnu \,\diff \nu
=
 \frac{\partial B }{\partial T} 
=
  \frac{\partial}{\partial T} \left(\frac{  \sigma_B T^{4}}{ \pi} \right) 
=
  \frac{4 \sigma_B T^{3}}{\pi}
\, ,
\label{eq:dbdtdef}
\eeq
where $\sigma_B$ is the Stefan-Boltzmann constant.
One can then write the frequency-integrated flux as a \emph{radiative diffusion equation},
\beq
F(r) = - \left [ 
\frac{16 \sigma_B}{3} \frac{T^{3}}{\kappa_{R} \rho} 
\right ] \,  
\frac{dT}{dr}
\ .
\label{eq:raddiff}
\eeq
The next section examines the stellar structure scalings for such radiative envelopes, both in terms of the classical ESM M-L scaling, and for detailed opacity models based on the OPAL tables.

\section{Effect of Radiation Pressure on  Stellar Envelope} 
\label{sec:envelope}
\label{sec3}

\subsection{Mass-Luminosity Scaling for Radiative Envelope}
\label{sec:M-L}
\label{sec3.1}

This diffusion form for energy transport, along with the requirement for momentum balance through hydrostatic equilibrium,  provide the basic stellar structure constraints that set the mass-luminosity scaling.
To see this, let us rewrite equation (\ref{eq:raddiff}) 
in terms of the radial gradient of the \emph{radiation} pressure $\Prad  \equiv 4 \sigma_B T^4/3c$,
\beq
\frac{d\Prad}{dr} = -  \rho  \frac{ \kappa_R F}{c} = - \rho g_{rad}  = - \rho \Gamma \frac{G\Mstar}{r^2}
\, .
\label{eq:dpraddr}
\eeq
When modified to account for the $1-\Gamma$ radiative reduction in the effective gravity,
the requirement for hydrostatic equilibrium sets the gradient of the \emph{gas} pressure,
\beq
 \frac{d\Pgas}{dr} = - \rho \frac{G\Mstar }{r^{2}} (1-\Gamma)
\, .
\label{eq:dpgasdr}
\eeq
Together these imply that the relative variation of gas to radiation pressure depends only on the Eddington parameter,
\beq
\frac{d\Pgas}{d\Prad} = \frac{1-\Gamma}{\Gamma}  
\, .
\label{eq:dpgdpr}
\eeq
As noted in \S\ref{sec:background}, for the Eddington Standard Model with constant $\Gamma$, this gives 
$\Pgas/\Prad \approx (1 - \Gamma)/\Gamma$= constant,  which 
leads to the simple ESM scaling (\ref{eq:esmgm}) for the mass dependence of the Eddington parameter.
A related, commonly quoted quantity is the gas pressure fraction $\beta$ of the total pressure, which in the interior  of an ESM model is just set by the Eddington parameter,
\beq
\beta \equiv \frac{\Pgas}{\Pgas + \Prad} = 1- \Gamma \, .
\label{eq:beta-gamma}
\eeq

These ESM scalings can be understood from average gradients in (\ref{eq:dpraddr}) and (\ref{eq:dpgasdr}) in terms of stellar mass $\Mstar$ and radius $R$.  Using the ideal gas law $\Pgas \sim \rho T$ with $\rho \sim \Mstar/R^3$,  (\ref{eq:dpgasdr})  implies the characteristic  interior temperature scales as
\beq
T 
\sim \frac {\Mstar 
(1-\Gamma)}{R}
\, .
\label{eq:tmbr}
\eeq
With the further proportionalities $F \sim \Lstar/R^2$ and $\Prad \sim T^4$, the radiative diffusion  (\ref{eq:dpraddr}) gives
\beq
\Lstar \sim \frac{R^{4} T^{4}}{\Mstar}  
\, .
\label{eq:lrtm}
\eeq
Combining (\ref{eq:tmbr}) and (\ref{eq:lrtm}), we can eliminate both $R$ and $T$ to find
\beq
\Lstar \sim 
(1-\Gamma)^{4} 
\,  \Mstar^{3} 
~~~~ {\rm or} ~~~ 
\frac{\Gamma}{(1-\Gamma)^4}  \sim \Mstar^2
\, ,
\label{eq:mlumscl}
\eeq
which agrees with the ESM scaling (\ref{eq:esmgm}).
As noted, this scaling does not depend explicitly on the nature of energy generation in the stellar core, 
but is strictly a property of the envelope structure\footnote{Of course, this simple scaling relation has to be modified to accommodate gradients in the molecular weight as a star evolves from the zero-age main sequence, and it breaks down altogether in the coolest stars (both giants and dwarfs), for which convection dominates the envelope energy transport.}.

\subsection{Virial Theorem and Stellar Binding Energy}
\label{sec3.2}

This hydrostatic balance of a stellar envelope can also be used to derive a relation --
known as the {\em virial theorem} -- between the internal thermal
energy $U$ and the gravitational binding energy $\Phi$ of the whole star \citep{Kippenhahn13},
\beq
\Phi  = - 3 (\gamma - 1) U
\, .
\label{eq:virial1}
\eeq
where $\gamma$ is the ratio of specific heats.
The total stellar energy is thus given by
\beq
E \equiv \Phi + U 
= \frac{3\gamma - 4}{3 \gamma - 3} \, \Phi 
\, .
\label{eq:totevir}
\eeq
For the case of a monotonic ideal gas $\gamma=5/3$, the total energy is the just half the gravitational binding energy,
$E = \Phi/2$.

However, in very massive stars near the Eddington limit $\Gamma \rightarrow 1$, the internal energy can 
become dominated by radiation instead of gas,  since $\Prad/\Pgas = \Gamma/(1-\Gamma) \rightarrow \infty$.
In this limit of a radiation gas, $\gamma \rightarrow 4/3$, which
by eqn.\ (\ref{eq:totevir}) implies a {\em vanishing} total energy $E \rightarrow 0$.
This is another factor toward making VMS unstable.

\subsection{OPAL opacity }
\label{sec:opal}
\label{sec3.3}

Let us next examine how the Rosseland opacity $\kapR$, and its associated Eddington parameter $\Gamma$, can change through the stellar envelope due to changes in temperature and density.
For this, we adopt the widely used tabulations from the OPAL\footnote{\url{http://opalopacity.llnl.gov/}} opacity project 
\citep{Iglesias96},
using the specific OPAL tables given by \citet{Grevesse93}, and taking the case with standard solar values $X=0.7$ and $Z=0.02$ for the Hydrogen and metal mass fractions.
The OPAL tabulations are given in terms of temperature $T$ and a parameter ${\cal R}  \equiv \rho/(T/10 ^6 K)^3$, but to make a clear connection to the above discussion, let us here cast the latter in the equivalent terms of gas to radiation pressure, $\Pgas/\Prad$.

\begin{figure}[t]
\begin{center}
\includegraphics[scale=0.46]{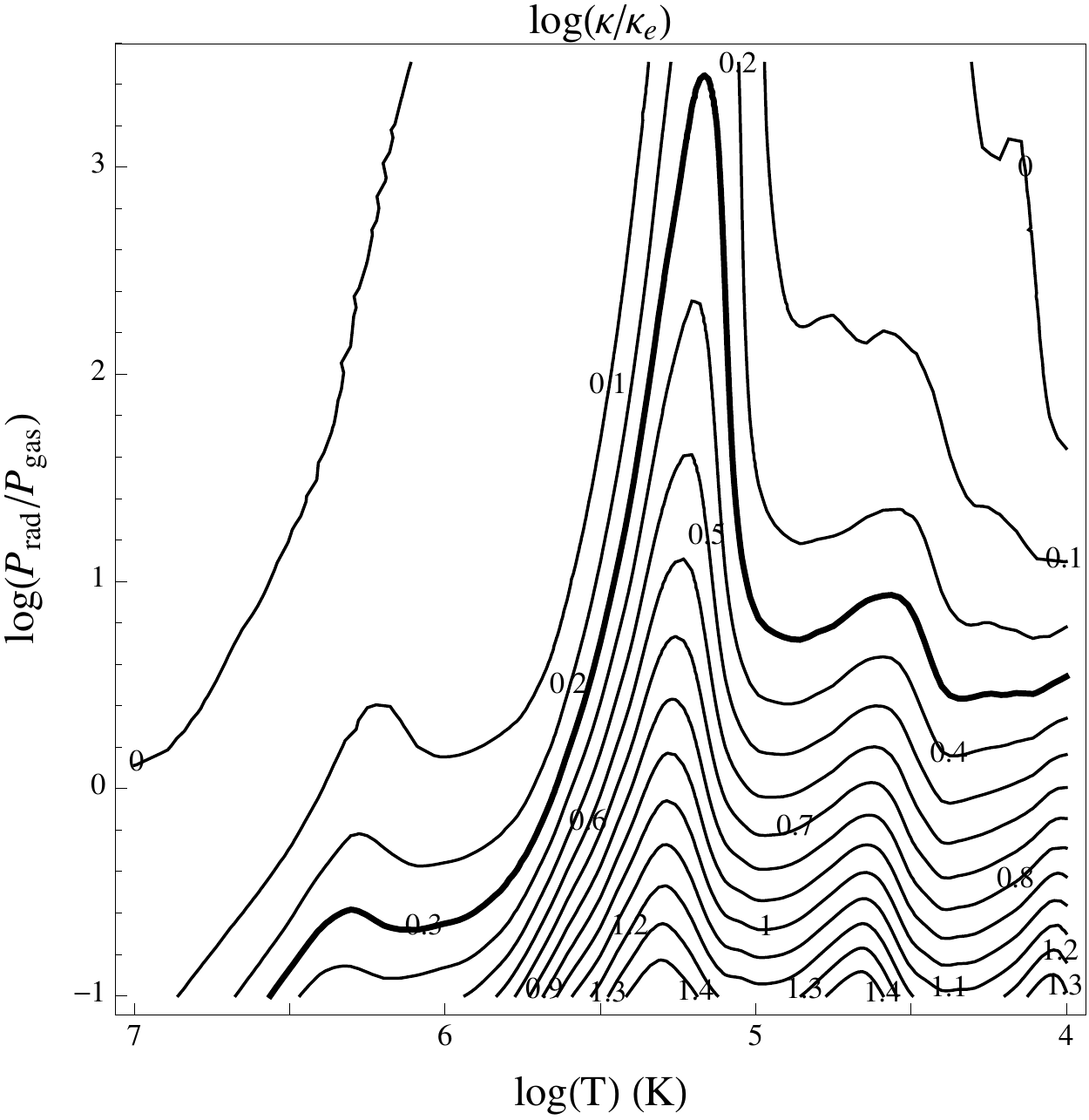}
\includegraphics[scale=0.45]{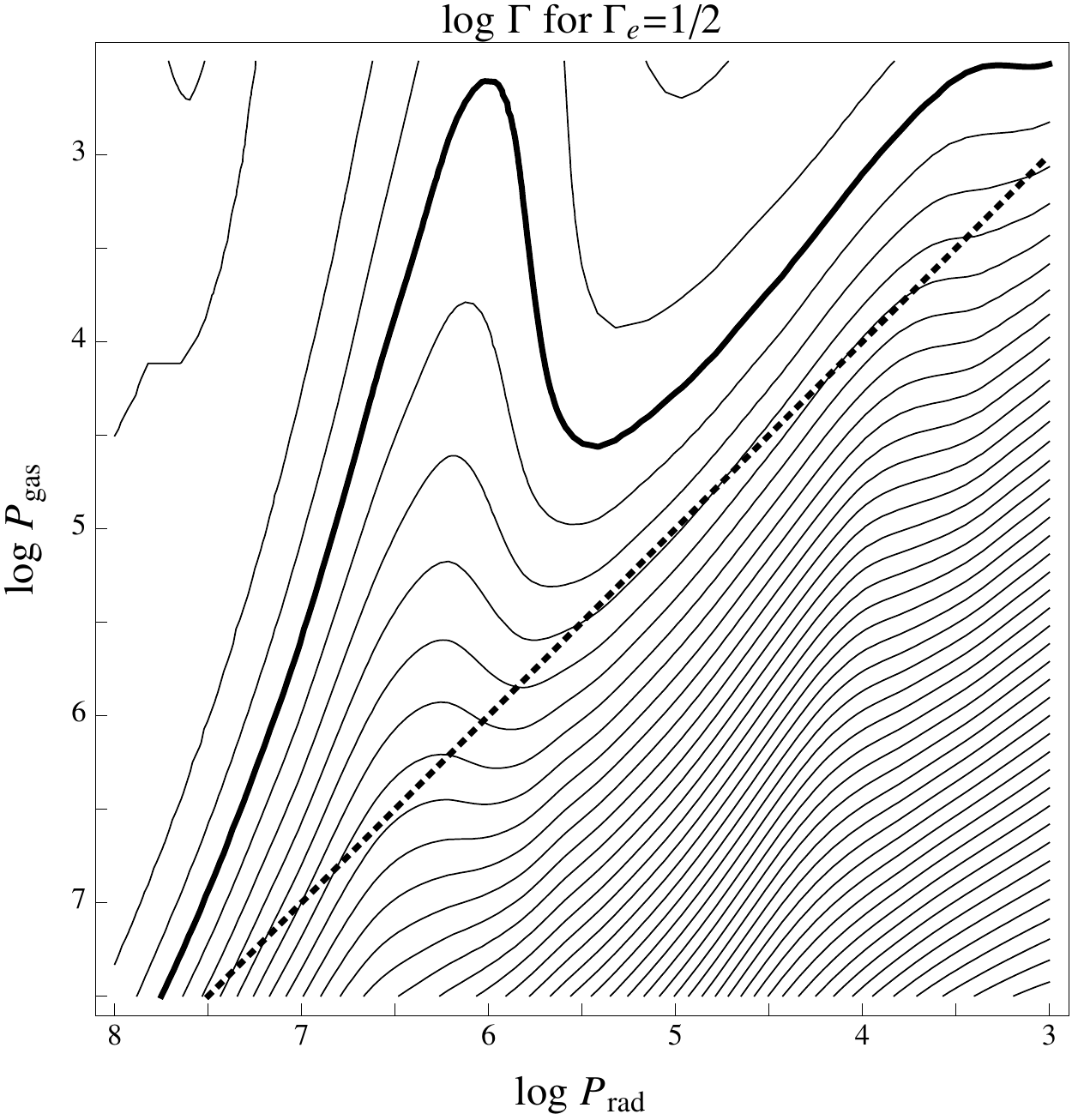}
\caption{
{\em Left:}
OPAL opacity $\kappa$, from tables by \citet{Grevesse93} for mass fractions $X=0.7$ and $Z=0.02$, plotted as contours of $\log (\kappa/\kappa_e)$ in the $\log T$ vs.\ $\log(\Prad/P_{gas)}$ plane.
The peak in contours at $\log T \approx $5.3 is from the Iron opacity bump.
{\em Right}: 
For a star with $\Gamma_e=1/2$, contours of $\Gamma = (\kappa/\kappa_e) \Gamma_e$, now plotted in the $\log \Prad$ vs.\ $\log \Pgas$ plane.
In both plots, the contours have log spacings of 0.1, with the heavy solid contours representing the case with $\kappa=2 \kappa_e$ and $\Gamma=1$. In the right panel, the dotted line shows the locus where $\Prad=\Pgas$.
}
\label{fig:opal}
\end{center}
\end{figure}

The left panel of figure \ref{fig:opal} plots contours of $\log(\kappa/\kappa_e)$ in the $\log(T)$ vs. $\log(\Prad/\Pgas)$ plane, oriented such that the high-temperature, high-density of the stellar interior is at the lower left.
The heavy black contour 
with labeled value 0.3 corresponds to an opacity $\kappa/\kappa_e \approx 2$ that is roughly twice that of the basal value for electron scattering. 
Indeed, for typical stellar-core temperatures of order several million Kelvin (MK), note that the total opacity
is only slightly above this electron scattering value.

But for temperatures near 1~MK and below, the opacity increases, especially for low values of $\Prad/\Pgas$ corresponding to relatively higher densities $\rho$.
Particularly note the strong peak near $T \approx 10^{5.3}$\,K $\approx 2 \times 10^5$\, K, which is due to a dense ``blanketing'' of line (bound-bound) opacity from iron group elements, and is thus commonly known as the ``{\em iron opacity bump'}'.
For low $\Prad/\Pgas$, and thus {\em high} density $\rho$, the opacity can exceed the electron scattering value by an order of magnitude or more.
This enhancement decreases at lower density, but only very weakly, requiring several decades decline in $\rho$ (represented here by a several decade increase in $\Prad/\Pgas$) to recover the modest factor 2 above electron scattering.

For temperatures along the opacity peak, an approximate fitting relation, normalized about the density $\rho = \rho_{2}$ that has $\kappa = 2 \kappa_e$, is given by the logarithmic form,
\beq
\frac{2 \kappa_e}{\kappa  } \approx  1+ \log \left ( \frac{\rho_2}{\rho} \right )^{0.2}  \, ,
\label{eq:kaplog}
\eeq
wherein the 0.2  quantifies the extreme weakness of the dependence on density.
For modest deviations about $\rho_{2}$, this can alternatively be written in the power-law form,
\beq
\frac{\kappa}{2 \kappa_e} \approx  \left ( \frac{\rho}{\rho_2} \right )^{0.086} \, ,
\label{eq:kappow}
\eeq
where the small power index $ 0.086 = 0.2 \log e $ again shows the weak density dependence.

\subsection{Envelope Inflation and the Iron Bump Eddington Limit}
\label{sec:ibel}
\label{sec3.4}

For VMS near the classical Eddington limit, this iron-bump increase in opacity near temperatures $T \approx 150,000-200,000$\,K, can have dramatic effects on the near-surface envelope structure, inducing an {\em inversion} in gas pressure and density, with an associated {\em inflation} in the surface radius \citep{Petrovic06, Graefener12},
and possibly triggering strong stellar wind mass loss.

For the specific case of a VMS with $\Gamma_e=1/2$, the right panel of figure \ref{fig:opal} plots contours of $\log \,\Gamma$ in the $\log \Prad$ vs.\ $\log \Pgas$ plane, with the heavy solid contour for the Eddington limit value $\Gamma=1$.
Since the combined stellar structure equation  (\ref{eq:dpgdpr}) specifies $d\Pgas/d\Prad$ in terms of $\Gamma$, we can follow the decline of gas pressure from  
 the deep interior,  where $\Gamma \approx \Gamma_e = 1/2$, implying from (\ref{eq:dpgdpr}) that  $\Pgas \approx \Prad$. But toward the subsurface layer with lower temperature and pressure, the large bump in opacity pushes the integration toward the $\Gamma =1$ contour, forcing the gas pressure, and thus density, to very low values, in this case to a minimum density $\rho_{min,2}  \approx 2.5 \times 10^{-11}$g\,cm$^{-3}$ at the peak of the bump, where the opacity is $\kappa = 2 \kappa_e$.

A key result here is that, because this iron-peak opacity depends so weakly on density, keeping it limited to the Eddington value $\kapedd \equiv \kappa_e/\Gamma_e$ requires a minimum density that has a {\em strong inverse} scaling with electron Eddington parameter. Using eqn.\ (\ref{eq:kaplog}), this can be fit approximately by
\beq
\log \left (\frac{\rho_{min,2}}{\rho_{min}} \right ) \approx 10( \Gamma_e  - 1/2 )
\, .
\label{eq:rhominge}
\eeq
Note, for example, that each linear increase of just 0.1 in $\Gamma_e$ gives an {\em order magnitude decrease} in $\rho_{min}$!

For a typical VMS effective temperature $\Teff \sim$60,000\,K, the optical depth at the Iron bump temperature $T \sim $180,000\,K is, by the diffusion equation (\ref{eq:dpraddr}), $\tau \sim (T/\Teff)^4 \sim 100$.
Because the iron bump region has very low density, and an opacity limited to the Eddington value $\kapedd = \kappa_e/\Gamma_e$, achieving this large $\tau \equiv \int \kappa \rho dr $ requires an extended, or ``inflated'', range in radius $r$.
Since $\Gamma \approx 1$, we see from the radiative diffusion equation (\ref{eq:dpraddr}) that the change in radiation pressures  scales as $d\Prad /\rho = G\Mstar d(1/r)$.
Integration from a ``core" radius $R_c$ at the base of the iron bump to an outer ``envelope" radius $R_e$ gives 
\beq
\frac{\Delta P}{2 \rho_{min}} \approx G\Mstar \left (\frac{1}{R_c} - \frac{1}{R_e} \right )
\, ,
\label{eq:trapint}
\eeq
 where $\Delta P \approx 2.2 \times 10^6$\,dyne\,cm$^{-2}$  characterizes the radiation pressure width of the iron  bump, and the factor two in the left-side denominator comes from simple trapezoidal rule integration.
Equation (\ref{eq:trapint})  can be readily solved to give  a simple analytic scaling for the radius inflation factor in terms of a dimensionless ratio $W$ of the pressure width to gravitational binding energy \citep{Graefener12},
 \beq
 \frac{R_e}{R_c} \approx \frac{1}{1-W}  ~~~~ ; ~~~~ W \equiv \frac{\Delta P}{2 \rho_{min} G\Mstar/R_c }
 = W_{1/2} 10^{10(\Gamma_e-1/2)} 
 \, .
 \label{eq:rerc}
 \eeq
Note that when $W$ approaches order unity, the envelope radius can become very large.

\begin{figure}[t]
\begin{center}
\includegraphics[scale=0.66]{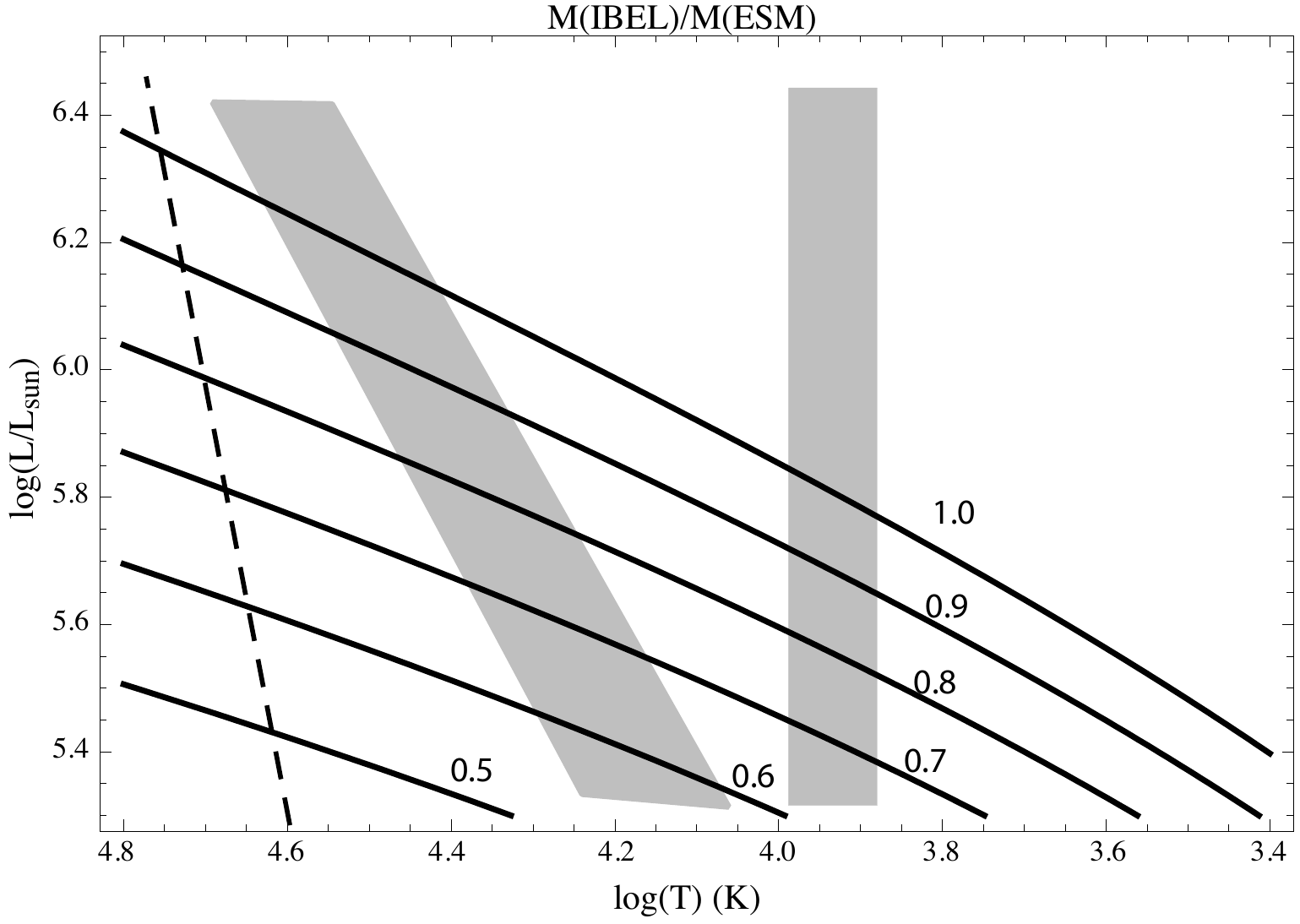}
\caption{
Uppermost part of H-R diagram, showing contours of the limiting mass for the iron bump Eddington limit, scaled by the mass for the Eddington standard model. 
The dashed curve shows the locus of the ZAMS, and the V-shaped shaded regions outline the observational range for S-Doradus LBVs.}
\label{fig:ibel}
\end{center}
\end{figure}

Indeed, the radius divergence $R_e \rightarrow \infty$ as $W \rightarrow 1$ defines a ``{\em Iron Bump Eddington Limit}" (IBEL), for which it is no longer possible to have a radiatively diffusive, hydrostatic envelope. Applying the scaling (\ref{eq:rhominge}) for $\rho_{min}$ within the critical condition $W=1$, we find the limiting Eddington parameter has the scaling
\beq
\Gamma_{IBEL} \equiv  \Gamma_e (W=1) \approx  0.5 + 0.1 \log \left ( \frac{\Mstar/\Msun}{13 R_c/\Rsun}  \right )
\, .
\label{eq:gamibel}
\eeq

To locate this limit on the H-R diagram, let us, for a given luminosity, associate the core radius with an effective temperature $T \sim \Lstar^{1/4}/R^{1/2}$. 
Figure \ref{fig:ibel} shows contours of the stellar mass for this limit, scaled for the given luminosity by the ESM mass of equation (\ref{eq:esmgm}),  and then plotted in an H-R diagram of $\log T$ vs.\ $\log(\Lstar/\Lsun)$.
Here the steep dashed curve represents the locus of the zero-age main sequence (ZAMS), and the  shaded regions outline the observational domains for S-Doradus type LBV stars, which typically vary horizontally within the V-shape on times scales of a few years.

Near the ZAMS, stars should roughly follow the ESM M-L scaling, 
and so the high location of the uppermost, unit-value contour representing the ESM mass indicates that only 
the most luminous VMS stars will breach the IBEL on the MS.
As less-massive MS stars evolve to the right, their luminosities tend to increase while the masses decrease due to mass loss, making $\Mstar/M_{ESM}$ decrease and so bringing them up against to the IBEL.
In principle, this could be an underlying cause or trigger for the observed variability in surface temperature of S-Doradus LBVs, but further work is needed to flesh out the nature of the time-dependent variations and relaxation processes.
As discussed by \citet{Petrovic06} and \citet{Graefener12},
the overall inflation effect may also  help explain the larger than expected core radii inferred for Wolf-Rayet stars.

A key general unresolved issue is the interplay between this iron-bump inflation and mass loss. Under what conditions might mass loss eliminate inflation, or inflation initiate mass loss?
Other complications include the potential roles of mixing, porosity, pulsation, etc. in limiting or disrupting the inflation effect from this idealized 1D, hydrostatic model.

 \section{Basic Formalism for Envelope Instability and Mass Loss}
 \label{sec4}

  \subsection{General  Time-Dependent Conservation Equations}
   \label{sec4.1}

The dominant role of radiative forces in VMS can lead to both time-dependent instabilities of the stellar interior and strong stellar wind mass loss from their surface.
To treat these we need to generalize the above assumption of a hydrostatic equilibrium balance to 
consider now cases 
wherein the vector sum of  forces acting on the gas is no longer zero, 
but instead has an imbalance that leads, via Newton's second law, to a net acceleration,
\beq
\frac{d \vvec}{dt} =  
\frac {\partial \vvec}{\partial t} +  \vvec  \cdot \nabla  \vvec = 
 \gvecrad
- g \rhat
- \frac{\nabla \Pgas }{ \rho } 
\, .
\label{tdeom}
\eeq
Here $\vvec$ is the flow velocity, 
and  $\gvecrad$ and $  -g \rhat $ and are the vector forms for the radiative acceleration and gravity,
with $\rhat$ a unit radial vector.
The first equality relates the total time derivative $d/dt$ as the sum of intrinsic variation $\partial/\partial t$ and advective changes along a flow gradient, ${\bf v} \cdot \nabla$.

The density and velocity are related through the mass conservation
relation
\beq
\frac{\partial \rho}{ \partial t} + \nabla \cdot  \rho {\bf v} 
= \frac{d \rho}{ d t} +  \rho \nabla \cdot  {\bf v} 
= 0 
\, ,
\label{tdmasscon}
\eeq
while conservation of energy takes the form
\beq
\frac{\partial e }{ \partial t} + \nabla \cdot  e{\bf v} = 
-\Pgas \, \nabla \cdot {\bf v} - \nabla \cdot \Fvec
\, ,
\label{tdeneqn}
\eeq
where the  divergence of vector radiative flux $\Fvec$
represents a local source or sink of gas internal energy $e$.
For an ideal gas with ratio of specific heats $\gamma$, this
is related to the gas pressure through
\beq
\Pgas = \rho a^{2} = (\gamma - 1)  e
\, ,
\label{pegm1e}
\eeq
where $a \equiv \sqrt{kT/\mu}$ is the isothermal sound speed, 
with $k$ Boltzmann's constant and $\mu$ the mean molecular weight.

Collectively,
eqns. (\ref{tdeom})-(\ref{pegm1e}) represent the general equations for 
a potentially time-dependent, multi-dimensional flow.

 \subsection{Local Linear Analysis for  ``Strange-Mode'' Instability of Hydrostatic Envelope}
 \label{sec:smodes}
  \label{sec4.2}
 
Let us first apply these general equations as the basis for a local, linear perturbation analysis of hydrostatic radiative envelopes in VMS with a dynamically significant radiative acceleration $\grad$.
The response of this acceleration to local perturbations, e.g.\ in the gas density $\rho$, gives rise to the so-called {\em strange-mode}  instability \citep{Blaes03, Glatzel93, Glatzel94, Glatzel05}.
The most general form requires a global analysis of the envelope structure;
but in the limit of wavelengths shorter than the local gravitational scale height, we can apply the so-called ``WKB approximation'' that ignores background gradients and analyzes the effect of localized, small-amplitude perturbations $\delta \rho$,  $\delta v$, etc.
(Since the coupling to radiation keeps the gas nearly isothermal, a simplified analysis can ignore perturbations in temperature.)

For strange modes, the simplest case involves purely radial variations of sinusoidal form  $\delta \sim e^{i(kr -\omega)}$, with $k$ the (real) radial wavenumber and $\omega$ the (possibly complex) frequency.
To first order in small-amplitude perturbations, the momentum equation (\ref{tdeom}) gives
 \beq
- i \omega \delta v = i k a^2 \frac{\delta \rho}{\rho} + \delta \grad
\, .
\label{eq:pertmom}
\eeq
A similar application to  the perturbed continuity equation (\ref{tdmasscon}) relates the perturbed velocity to density,
\beq
{\delta v } = \frac{\omega}{k} \, \frac{\delta \rho}{\rho} 
\, .
\label{eq:pertcont}
\eeq
If we further assume that the opacity has density dependence given by the logarithmic derivative
\beq
\Theta_\rho \equiv \frac{\partial \ln \kappa}{\partial \ln \rho}
\, ,
\label{eq:thetarho}
\eeq
then the perturbed radiative acceleration can also be expressed in terms of the perturbed density,
\beq
\delta \grad = \Theta_\rho \grad \frac{\delta \rho}{\rho}
\, .
\label{eq:dgradrho}
\eeq
The combination of (\ref{eq:pertmom}), (\ref{eq:pertcont}), and (\ref{eq:dgradrho}) allows us to solve for a {\em dispersion relation} for the frequency
\beq
\omega = \sqrt{ a^2 k^2 + i k \Theta_\rho \, \grad} 
\approx \pm ak  \left ( 1 + \frac{i \Theta_\rho \grad}{2 a^2 k} \right )
= \pm ak  \pm i \Theta_\rho \frac{ \grad}{2 a} 
\, ,
\label{eq:omnod}
\eeq
where the middle approximation uses a high-wavenumber limit to expand the radical from the first form.
For upward-propagating (+) modes, the frequency has a positive imaginary component, implying an instability with growth rate  $\eta = \Theta_\rho \, \grad/2 a = \Theta_\rho \Gamma g/2a$.
The associated instability growth time is 
\beq
t_g \equiv \eta^{-1}
 = \frac{2a}{\Theta_\rho \Gamma g} 
= 400 \,{\rm s} \,
 \frac
 {a_{20}}
 {g_4 \Gamma \Theta_\rho}
\, ,
\eeq
where $a_{20} \equiv a/(20 {\rm km/s})$ and $g_4 \equiv g/(10^4 {\rm cm/s^2})$.

Figure \ref{fig:smode}a shows that for OPAL opacities $\Theta_\rho$ is typically a few tenths in the subsurface regions of a stellar envelope.
Taking $ \Theta_\rho \approx 0.5$ and typical stellar parameters $g_4 = a_{20} = 1$,  we see that instability growth time $t_g \sim  $\,2000\,s is quite short for VMS with $\Gamma $ a factor few below unity, but is  much longer for lower-mass stars with very small $\Gamma$.

\begin{figure}
\includegraphics[width=5.5cm]{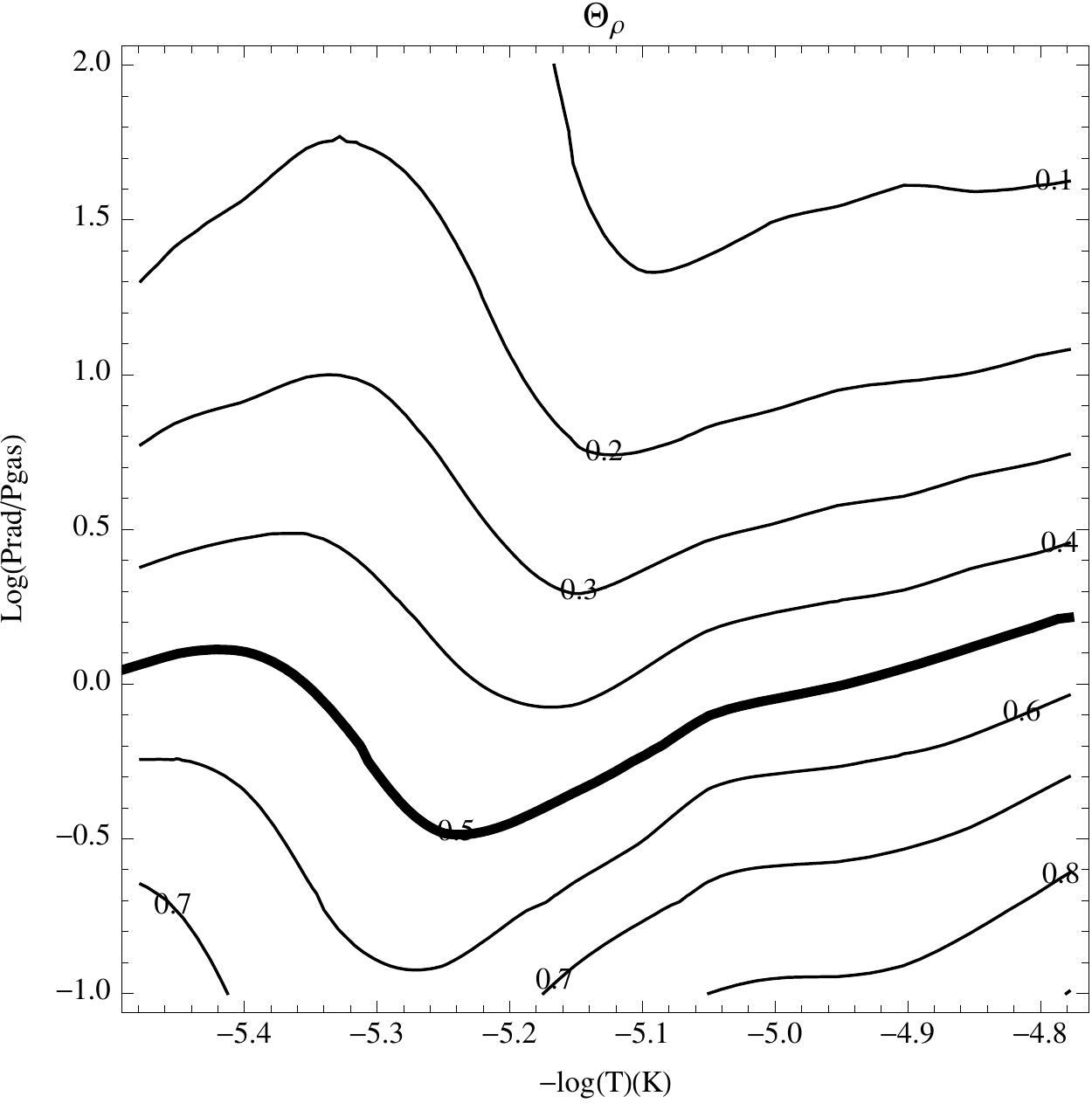}
\includegraphics[width=5.5cm]{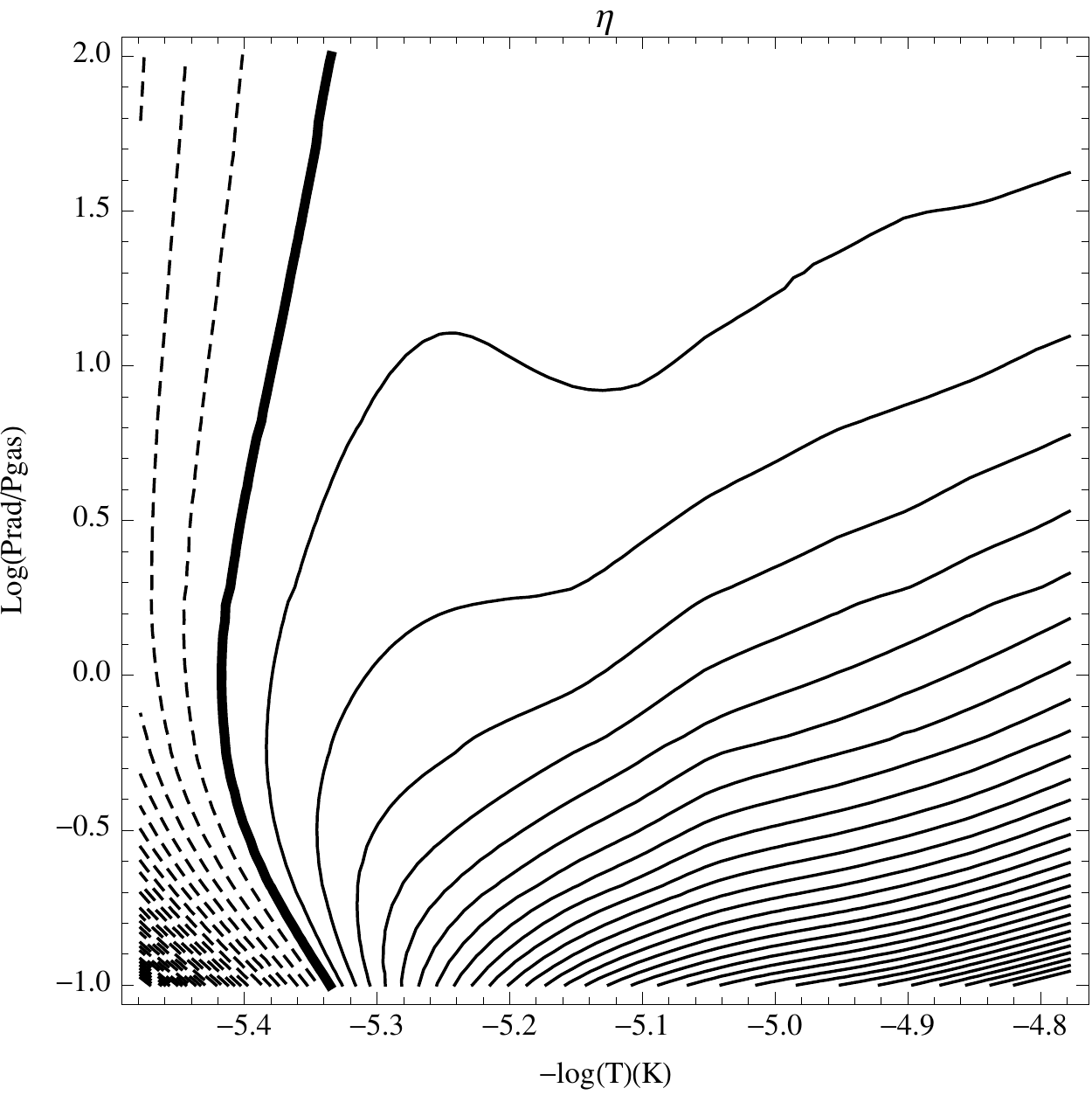}
\caption{
Logarithmic derivative of opacity with density $\Theta_\rho$ (left) and 
net instability growth rate $\eta$ (scaled by $\grad/2a$; right) 
from eqn.\ (\ref{eq:etasm}), 
plotted as contours in the $\log T$ vs. $\log (\Prad/\Pgas ) $ 
plane for the case of a hot VMS with 
$\Teff = 60kK$.x
($\log \Teff =  4.8$).
The contours are in increments of 0.1 and the heavy thick curves represent a value of 0.5 for $\Theta_\rho$ and 0 for $\eta$. The dashed contours indicate negative growth rates, and so show deep interior regions have a net damping due to radiation drag effects.
}
\label{fig:smode}
\end{figure}

The above stability analysis is based solely on mass and momentum conservation, ignoring the gas and radiative energy.
A generalization 
\citep{Blaes03} that takes proper account of associated perturbations in the radiation field shows that the instability competes against the damping from a ``radiation drag'' effect, yielding now a net growth rate
(cf.\ eqn.\ 62 of \citet{Blaes03})
\beq
\eta = \frac{\grad}{2 a} \left ( 1 + \frac{\Pgas}{4 \Prad} \right ) ( \Theta_\rho - D )
\, ,
\label{eq:etasm}
\eeq
where 
the radiative drag coefficient term is given by
\beq
 D 
 = \left ( 1 + \frac{\Pgas}{4 \Prad} \right ) \frac{4\Prad}{F} a 
 = \left ( 1 + \frac{\Pgas}{4 \Prad} \right ) \frac{16}{3} \left ( \frac{T}{\Teff} \right )^4 \, \frac{a}{c} \, .
 \eeq
As shown in the right panel of figure \ref{fig:smode},
in the hot interior  regions with $\log T \gtrsim 5.4$, the medium is now stabilized by the net damping from this radiation drag term,
but 
in the subsurface layers where $T \lesssim \Teff$, the $a/c$ factor keeps this drag small, thus preserving the instability found above, and so again giving growth rates $\eta$ that are a few tenths times $\grad/2a$.

The net instability should lead to amplification of upward propagating sound waves,  eventually limited by the steeping into weak or moderate shocks with velocity amplitude on the order of the sound speed, $\delta v \sim a$.
For near-surface temperatures $T \sim \Teff$ of a few $10^4$\,K, the sound speed is $a \sim 20$\,km/s, much smaller (by factor $\sim$1/30) than the near-surface escape speed  $\vesc \equiv \sqrt{2 GM/R}$, which is about 600\,km/s for a star with a solar value for the ratio of mass to radius.
As such, this short-wavelength form of such strange mode instability is not at all suitable to providing the kind of large-scale mass ejection inferred from LBVs.  
But there have been suggestions \citep{ Glatzel93, Glatzel94, Glatzel05} that analogous larger-scale, global modes of strange-mode pulsations might be important in triggering episodic mass loss.

\subsection{General Equations for Steady, Spherically Symmetric Wind}
 \label{sec4.3}

The general flow conservation equations (\ref{tdeom})-(\ref{pegm1e})  also provide the basis for modeling stellar wind outflows.
First-order wind models are commonly based on the 
simplifying approximations of steady-state ($\partial/\partial t=0$), 
spherically symmetric,
radial outflow (${\bf v} = v (r) {\bf \hat  r}$).
The mass conservation requirement (\ref{tdmasscon}) then can then be used 
to define a constant overall mass loss rate,
\beq
{\dot M} \equiv 4 \pi \rho v r^{2} 
\, .
\label{mdotdef}
\eeq
Using this and the ideal gas law (\ref{pegm1e}) to eliminate the density in the 
pressure gradient term then gives for the radial equation of motion
\beq
\left ( 1 - 
\frac{ a^{2}
}{
v^{2} }
\right ) 
v \, \frac{dv }{ dr} =
 \grad 
 -  \frac { G \Mstar }{ r^{2}}
+ \frac{ 2 a^{2} }{ r } 
+ \frac{d a^{2} }{ dr }
\, .
\label{ssradeom}
\eeq

The gas pressure terms (containing the sound speed  $a$) on the right-hand-side are key to accelerating the hot (MK) coronal-type winds of the sun and other cool, lower-mass stars. 
But in winds from massive stars -- which are kept almost isothermal near the stellar effective temperature by the competition between photoionization heating and radiative cooling -- these terms are negligible, since compared to competing terms  needed to drive the wind, they are of order  $\ws \equiv (a/\vesc)^{2} \approx 0.001$, where  $\vesc \equiv \sqrt{2G\Mstar/\Rstar}$  is the  escape speed from the stellar surface radius $\Rstar$.
These gas pressure terms on the right-hand side of (\ref{ssradeom}) can thus be quite generally neglected in VMS winds.
However, to allow for a smooth mapping of  a wind model onto a hydrostatic atmosphere through a subsonic wind base, one can still retain the sound-speed term  on the left-hand-side.
Transitioning to a supersonic wind  then requires $\grad = GM/r^2$, and so $\Gamma=1$, at the sonic point $v=a$.

Since overcoming gravity is key, it is convenient 
to rewrite (\ref{ssradeom}) in a dimensionless 
form that scales all  accelerations by gravity,
\beq
\left ( 1 - \frac{\ws}{w} \right )
\, w' = \Gamma - 1
\label{dimlesseom}
\eeq
Here 
$w \equiv v^{2}/\vesc^{2}$ is the flow kinetic energy in terms of the escape energy from the surface radius $\Rstar$,
and $w' \equiv dw/dx$ is the change of this  scaled energy with scaled the gravitational potential  $x \equiv 1-\Rstar/r$ at any radius $r$.

In characterizing the sonic point as the flow ``critical point'', it is sometimes suggested \citep{Nugis02} 
that conditions for reaching $\Gamma =1$, for example in the iron opacity bump, set the sonic point density $\rho_s$ and thus the mass loss rate $\Mdot = 4 \pi \rho_s a R^2$ of a steady wind outflow.
But it is important to emphasize that, because the flow energy at the sonic point is just a tiny fraction $\ws \approx 0.001$ of what's needed to escape the star's gravitational potential, maintaining a steady wind requires keeping $\Gamma > 1$ over an extended range of the supersonic region.
In \S \ref{sec6.3} we discuss the flow stagnation that occurs if the opacity or radiative energy flux is insufficient to maintain initial outflow from a limited super-Eddington region.
But the next section first reviews the standard ``CAK'' theory  \citep{CAK75} for a steady-state wind outflow driven by line-opacity.

\section{Line-Driven Stellar Winds}
 \label{sec5}

As noted in \S\S \ref{sec:planck} and \ref{sec:kapr}, the resonant nature of line (bound-bound) scattering from metal ions leads to an opacity that is inherently much stronger than from free electrons.
In the deep envelope where the radiative transfer is well characterized as a radiative diffusion, the cumulative opacity of lines is given by the Rosseland mean, which can be several times that from pure electron scattering, e.g.\ near the Iron bump;
in VMS approaching the IBEL, this can lead to a strong inflation of the stellar envelope (\S \ref{sec:ibel}).

But in the idealized case that ions are illuminated by an unattenuated continuum -- as would occur in the limit of {\em optically thin} radiative transfer--, the cumulative opacity is given by the {\em Planck} mean 
(\S \ref{sec:planck}), wherein the dominant contribution of lines leads to an opacity enhancement that is a huge factor ${\overline Q} \approx 2000 $ larger than from free electrons \citep{Gayley95}.
In terms of the radiative force, this implies the Eddington parameter associated with lines can be enhanced by a similarly large factor over the classical electron scattering value.
This suggest that, even in moderately massive stars with electron Eddington parameters $\Gamma_{e} > 5 \times 10^{-4}$, line opacity could completely overcome gravity and so initiate a sustained stellar wind outflow.

In practice, self-absorption within strong lines limits the line force, giving it a value that is intermediate between that from the Rosseland and Planck means.
Modeling such line-driven wind outflows thus requires a treatment of the line radiation transport in regimes between the associated diffusion vs. optically thin limits.
Chapter 4 reviews how mass loss rates from VMS are computed from numerical models using Monte-Carlo (MC) treatments of the radiation transport.
As complement to this, the next section (\ref{sec:caksob}) reviews the classical \citet[CAK]{CAK75} model, which by using the key Sobolev approximation \citep{Sobolev60} for localized line-transport, allows one to obtain fully {\em analytic} solutions for the steady-state wind, with associated simple scaling forms for the wind mass loss rate and velocity law
\citep{Owocki13}.
This provides a basis for a linear perturbation analysis of a strong instability intrinsic to line-driving 
(\S \ref{sec5.2.1}), and for time-dependent numerical hydrodynamical simulations of the resulting instability wind structure (\S \ref{sec5.2.2}).

\subsection{The CAK/Sobolev Model for Steady-State Winds}
\label{sec:caksob}
 \label{sec5.1}

\subsubsection{Sobolev line-transfer and desaturation by wind expansion}
 \label{sec5.1.1}


\begin{figure}
  \includegraphics[width=11.5cm]{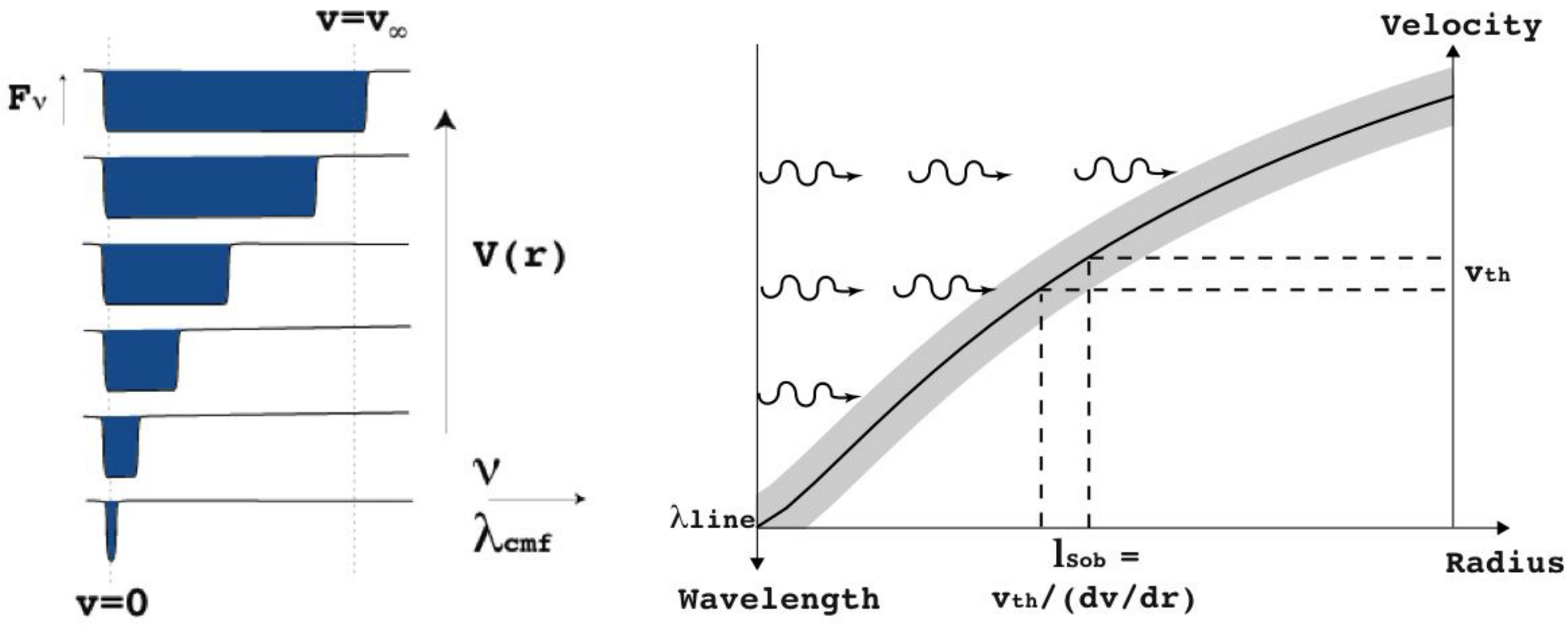}
\caption{
Two perspectives for 
the Doppler-shifted line-resonance in an accelerating flow. 
Right:
Photons with a wavelength just shortward of a line 
propagate freely from the stellar surface up to a layer
where the wind outflow Doppler shifts the line into a resonance over a 
narrow width (represented here by the shading) equal to the Sobolev length,
set by the ratio of thermal speed to velocity gradient, 
$l_{Sob} \equiv v_{th}/(dv/dr)$. 
Left: 
Seen from successively larger radii within the accelerating wind, 
the Doppler-shift sweeps out an increasingly broadened line
absorption trough in the stellar spectrum.
}
\label{fig:2dopps}
\end{figure}

As illustrated in figure \ref{fig:2dopps}, a key factor in controlling the net strength of the line-force that drives a stellar wind outflow is the desaturation of the lines associated with the variable Doppler shift from the wind acceleration.
In the highly supersonic wind,
 the thermal Doppler broadening of the line, which for heavy ions is set by a thermal speed $v_{th}$ that is a factor several smaller than the sound speed $a$, is much smaller than the Doppler shift associated with the wind outflow speed $v \gg a$.
This allows a localized ``Sobolev approximation" \citep{Sobolev60} for the line transport, with stellar photons interacting with the wind over a narrow resonance layer, with width set by the Sobolev length, $l_{Sob} = v_{th}/(dv/dr)$, 
and with associated optical depth proportional to
$t_{} \equiv \kappa_{e} \rho c /(dv/dr) $ 
$= \Gamma_{e} {\dot M} c^{2}/L_{\ast} w^{\prime}$.
%
%

\subsubsection{The CAK line-force, mass loss rate, and wind velocity law}
 \label{sec5.1.2}

Based on this Sobolev treatment of line transport,  \citet{CAK75} developed a powerful and highly useful formalism for treating the cumulative force from a large ensemble of lines by assuming they could be approximated by power-law number distribution in line-strength, with characteristic power index $\alpha$.
A key result is that the cumulative force from the ensemble is reduced
by a factor  $1/({\overline Q}t)^{\alpha}$ from the optically thin 
value,
\begin{equation}
\Gamma_{CAK}
= \frac{ {\overline Q} \Gamma_{e}}{(1-\alpha) ({\overline Q} t_{} )^{\alpha} } \, 
= \Gamma_{e} k t^{-\alpha} = C (w')^{\alpha} \, ,
\label {gamcak}
\end{equation}
where the second equality defines the CAK ``force multiplier''
$kt^{-\alpha}$, with\footnote{
Here we use a slight variation of the standard CAK notation in which
the artificial dependence on a fiducial ion thermal speed is avoided by
simply setting $v_{th} = c$.Œ
Back-conversion to CAK notation is achieved by multiplying
$t$ by $v_{th}/c$ and $k$ by $\left ( v_{th}/c \right)^{\alpha}$.
The line normalization ${\overline Q}$ 
offers the advantages of being a dimensionless measure of line-opacity 
that is independent of the assumed ion thermal
speed,  with a nearly constant characteristic value of order
${\overline Q} \sim 10^3$ for a wide range of ionization conditions
 \citep{Gayley95}.} 
$k \equiv {\overline Q}^{1-\alpha}/(1-\alpha)$.
The last equality relates the line-force to the flow acceleration,
with
\begin{equation}
C \equiv
\frac {1}{1-\alpha} \, \left [ \frac {L_{\ast}} {{\dot M} c^2} \right ]^\alpha \,
\left [ {\overline Q} \Gamma_{e} \right ]^{1-\alpha} \, .
\label {cdef}
\end{equation}
Note
that, for fixed sets of parameters for the star ($L_{\ast}$, $M_{\ast}$, 
$\Gamma_{e}$) and
line-opacity ($\alpha$, ${\overline Q}$), this constant scales with the mass loss rate 
as $C \propto 1/{\dot M}^{\alpha}$.

Neglecting the small sound-speed term $w_{s} \approx 0.001 \ll 1 $,
application of eqn.\ (\ref{gamcak}) into (\ref{dimlesseom}) gives 
the CAK equation of motion,
\begin{equation}
F = w' + 1 - \Gamma_{e} - C (w')^\alpha = 0 \, .
\label{cakeom}
\end{equation}
For small ${\dot M}$ (large $C$), there are two solutions, while for
large ${\dot M}$ (small $C$), there are no solutions.
The CAK critical solution corresponds to a {\it maximal} mass loss rate,
defined by $\partial F/\partial w' = 0$, for which the $C(w')^{\alpha}$ is
tangent to the line $1-\Gamma_{e} + w'$ at a critical acceleration
$w'_{c} = (1-\Gamma_{e}) \alpha/(1-\alpha )$.
Since the scaled equation of motion (\ref{cakeom}) has no explicit 
spatial dependence, this critical acceleration  applies throughout 
the wind, and so can be trivially integrated to yield 
$w(x) = w_{c}'\, x$.
In terms of dimensional quantities, this represents
a specific case of the general ``beta''-velocity-law,
\begin{equation}
v(r)=
v_\infty
\left ( 1- \frac {R_{\ast}}{r} \right )^{\beta} \, ,
\label{CAK-vlaw}
\end{equation}
where here $\beta=1/2$, and
the wind terminal speed $v_\infty = v_{esc} 
\sqrt{\alpha(1-\Gamma_{e})/(1-\alpha)}$.
Similarly, the critical value 
$C_{c}$
yields, through eqn.\ (\ref{cdef}), 
the standard CAK scaling for the mass loss rate
\begin{equation}
{\dot M}_{CAK}=\frac {L_{\ast}}{c^2} \; \frac {\alpha}{1-\alpha} \;
{\left[ \frac {{\overline Q} \Gamma_{e}}{1- \Gamma_{e}} \right]}^{{(1-\alpha)}/{\alpha}} \, .
\label{mdcak}
\end{equation}

\subsubsection{Modifications and limitations of the CAK mass loss scaling}
\label{sec:caklimits}
 \label{sec5.1.3}

These CAK results strictly apply only under the idealized assumption that
the stellar radiation is radially streaming from a point-source.
If one takes into account the finite angular extent of the stellar disk, 
then near the stellar surface the radiative force is reduced by 
a factor $f_{d\ast} \approx 1/(1+\alpha)$,
leading to a reduced mass loss rate 
 \citep{FA86, PPK86}
\begin{equation}
{\dot M}_{fd} = f_{d\ast}^{1/\alpha} {\dot M}_{CAK} 
= \frac{ {\dot M}_{CAK}}{(1+\alpha)^{1/\alpha}}
 \approx {\dot M}_{CAK}/2 \, .
\label{mdfd}
\end{equation}
Away from the star, the correction factor increases back toward unity,
which for the reduced base mass flux implies a stronger, more extended
acceleration, giving a somewhat higher terminal speed,
$v_{\infty} \approx 3 v_{esc}$, and a flatter velocity law,
approximated by replacing the exponent in
eqn.\ (\ref{CAK-vlaw}) by $\beta \approx 0.8$.

The effect of a radial change in ionization can be approximately taken into account
by correcting the CAK force (\ref{gamcak}) by a factor of the form
$
\left ( {n_e / W } \right )^\delta ,
$
where $n_e$ is the electron density, 
$W \equiv 0.5 ( 1-\sqrt{1-R_{\ast}/r} )$ 
is the radiation ``dilution factor'', and the exponent has a typical value 
$\delta \approx 0.1$ 
\citep{Abbott82}.
This factor introduces an additional density dependence to that already implied
by the optical depth factor $1/t_{}^\alpha$ given in eqn.\
(\ref{gamcak}).
Its overall effect can be roughly accounted with the simple
substition $\alpha \rightarrow \alpha' \equiv \alpha - \delta$ in the power
exponents of the CAK mass loss scaling law (\ref{mdcak}).
The general tendency is to moderately increase ${\dot M}$, and accordingly to
somewhat decrease the wind speed.

The above scalings also ignore the finite gas pressure associated with a
small but non-zero sound-speed parameter $w_{s}$.
Through a perturbation expansion of the equation of motion
(\ref{dimlesseom}) in this small parameter, it possible
to derive simple scalings for the fractional corrections 
to the mass loss rate and terminal speed
 \citep{OuD04}
\begin{equation}
\delta m_{s} \approx \frac{ 4\sqrt{1-\alpha}}{ \alpha} \, \frac{a }{ v_{esc}}
~~~~ ;  ~~~~ \delta v_{\infty,s} 
\approx \frac{ - \alpha \delta m_{s} }{ 2(1-\alpha) }
\approx \frac{-2 }{ \sqrt{1-\alpha} } \, \frac{a }{ v_{esc} } \, .
\label{dmdvs}
\end{equation}
For a typical case with $\alpha \approx 2/3$ and $w_{s}=0.001$, 
the net effect is to increase the mass loss rate and decrease the wind 
terminal speed, both by about 10\%.

An important success of these CAK scaling laws is the theoretical
rationale
they provide for an empirically observed ``Wind-Momentum-Luminosity'' 
(WML) relation 
\citep{Kudritzki99}.
Combining the CAK mass-loss law (\ref{mdcak}) together with the 
scaling of the terminal speed 
with the effective escape,
we obtain a WML relation of the form,
\begin{equation}
    {\dot M} v_{\infty} \sqrt{R_{\ast}} \sim L^{1/\alpha'} 
{\overline Q}^{1/\alpha'-1} 
\end{equation}
wherein we have neglected a residual dependence on $M(1-\Gamma_{e})$ that
is generally very weak for the usual c ase that $\alpha'$ is near $2/3$. 
Note that the direct dependence ${\overline Q} \sim Z$ provides the
scaling of the WML with metalicity $Z$.

Finally, as a star approaches the classical Eddington limit $\Gamma_{e} 
\rightarrow 1 $, these standard CAK scalings formally predict the mass loss rate 
to diverge as ${\dot M} \propto 1/(1-\Gamma_{e})^{(1-\alpha)/\alpha}$, but 
with a vanishing terminal flow speed $v_{\infty} \propto \sqrt{1-\Gamma_{e}}$.
The former might appear to provide an explanation for the large mass 
losses inferred in LBV's, but the latter fails to explain the 
moderately high inferred ejection speeds, e.g. the  500-800 km/s 
kinematic expansion inferred for the Homunculus nebula of $\eta$~Carinae
\citep{Smith02, Smith03}.

So one essential point is that line-driving could never explain the
extremely large mass loss rates needed to explain the Homunculus
nebula.
To maintain the moderately high terminal speeds, the
$\Gamma_{e}/(1-\Gamma_{e})$ factor would have to be of order unity.
Then for optimal realistic values $\alpha=1/2$ and $Q \approx 2000$
for the line opacity parameters 
\citep{Gayley95},
the maximum mass loss from line driving is given by
\citet{SO06},
\begin{equation}
{\dot M_{max,lines}} \approx 1.4 \times 10^{-4} L_{6} \, M_{\odot}/yr
\, ,
\end{equation}
where $L_{6} \equiv L/10^{6} L_{\odot}$. 
Even for peak luminosities of a few times $10^{7} L_{\odot}$
during $\eta$~Carinae's eruption, this limit is still several
orders of magnitude below the mass loss needed to form the Homunculus.
Thus, if mass loss during these eruptions occurs via a wind,
it must be a super-Eddington wind driven by continuum radiation
force (e.g., electron scattering opacity) and not lines
\citep{Quinn85, Belyanin99}.
Such continuum-driven wind models for LBVs are discussed further in \S \ref{sec6}.

\subsection{Non-Sobolev Models of Wind Instability}
 \label{sec5.2}

The above CAK steady-state model depends crucially on the use of the
Sobolev approximation to compute the local CAK line force
(\ref{gamcak}).
Analyses that relax this approximation show that the flow is subject
to a strong, ``line-deshadowing instability'' (LDI) for 
velocity perturbations on a scale near and
below the Sobolev length $l_{Sob} = v_{th}/(dv/dr)$
 \citep{Lucy70, MacGregor79, OR84, OR85, OP96}.
Moreover, the diffuse, scattered component of the line force, 
which in the Sobolev limit is nullified by the fore-aft symmetry of
the Sobolev escape probability (see figure \ref{fig2}), 
turns out to have important dynamical effects on  the instability
through a ``diffuse line-drag"
\citep{Lucy84}.

\subsubsection{Linear Analysis of Line-Deshadowing Instability}
 \label{sec5.2.1}

For sinusoidal perturbations ($\sim e^{i(kr-wt)}$) with wavenumber 
$k$ and frequency $\omega$, the linearized momentum equation (\ref{eq:pertmom})
(ignoring the small gas pressure by setting $a=0$) relating
the perturbations in velocity and radiative acceleration implies
$
\omega = i \frac{\delta g}{\delta v}
$,
which shows that unstable growth, with $\Im{\omega} > 0$, requires 
$\Re{(\delta g/\delta v}) > 0$.
For a purely Sobolev model 
\citep{Abbott80},
the CAK scaling of the
line-force (\ref{gamcak}) with velocity gradient $v'$ implies
$\delta g \sim \delta v' \sim ik \delta v$, giving a purely real
$\omega$, and thus a stable wave that propagates inward at phase
speed,
\begin{equation}
\frac{\omega}{k} = - \frac{\partial g}{\partial v'} 
\equiv - U 
\, ,
\label{dgcak}
\end{equation}
which is now known as the ``Abbott speed''.
\citet{Abbott80} showed this is comparable to the outward wind flow speed, and
in fact exactly equals it at the CAK critical point.

\begin{figure}
\includegraphics[width=11.5cm]{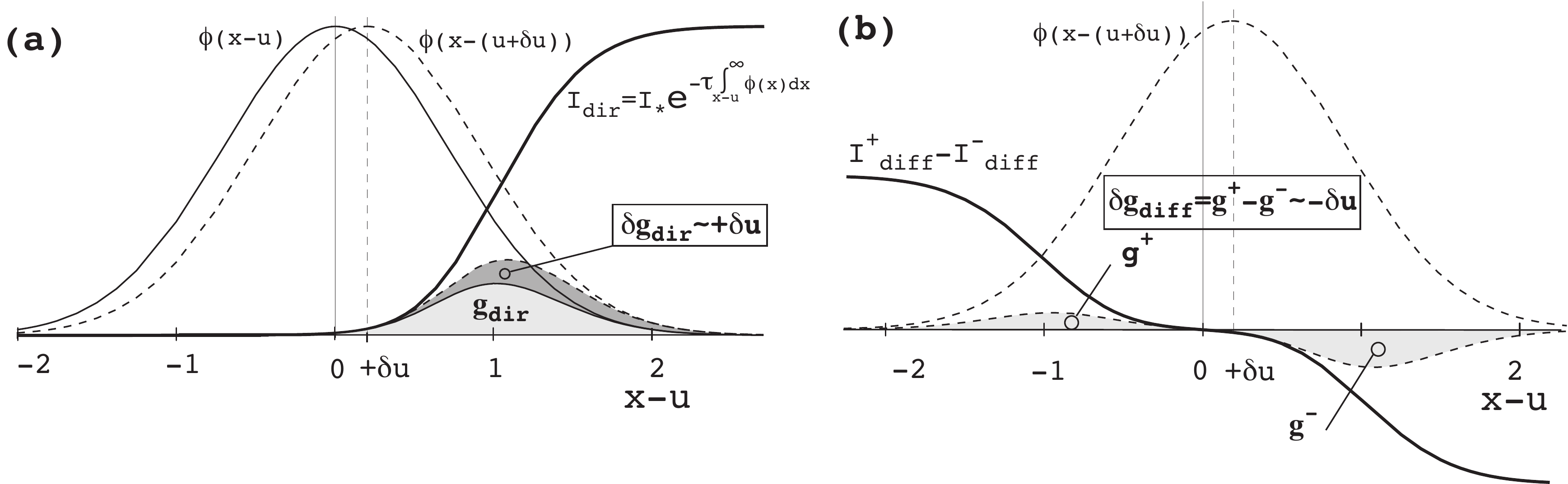}
\caption{
(a) The line profile $\phi$ and direct intensity plotted vs.
comoving frame frequency $x-u = x-v/v_{th}$, 
with the light shaded overlap area proportional 
to the net direct line-force $g_{dir}$.
The dashed profile shows the effect of the Doppler shift from a 
perturbed velocity $\delta v$, with 
the resulting extra area in the overlap with the blue-edge intensity 
giving a perturbed line-force $\delta g $ that scales in proportion to this
perturbed velocity $ \delta u = \delta v/v_{th}$.
(b) 
The comoving-frequency variation of the forward (+) 
and backward (-) streaming parts of the diffuse, scattered radiation.
Because of the Doppler shift from the perturbed velocity, 
the dashed profile has a stronger interaction with the backward 
streaming diffuse radiation, resulting in a diffuse-line-drag force 
that scales with the negative of the perturbed velocity, 
and so tends to counter the instability of the direct line-force in part a.
}
\label{fig2}
\end{figure}

As illustrated in figure \ref{fig2}a, instability arises from the deshadowing of
the line by the extra Doppler shift from the velocity perturbation,
giving $\delta g \sim \delta v$ and thus $\Im{\omega} > 0$.
A general analysis 
\citep{OR84} 
yields a ``bridging law'' encompassing both
effects,
\begin{equation}
\frac{\delta g_{} }{ \delta v} 
\approx \Omega \frac { i k \Lambda }{  1 + ik \Lambda }
\, ,
\label{dgbridge}
\end{equation}
where 
$\Omega \approx g_{cak}/v_{th}$
sets the instability growth rate,
and the ``bridging length'' $\Lambda$ is found to be of order the 
Sobolev length $l_{sob}$.
As illustrated in figure \ref{fig:abbldi}, in the long-wavelength limit $k \Lambda \ll 1$,
we recover the stable, Abbott-wave 
scalings of the Sobolev approximation,
$\delta g_{}/ \delta v  \approx i k \Omega \Lambda = ik U$;
while in the 
short-wavelength limit $k \Lambda \gg 1$,
we obtain the instability scaling
$\delta g_{}  \approx \Omega \delta v$.
The instability growth rate is very large, about the flow 
rate through the Sobolev length, $\Omega \approx v/l_{Sob}$.
Since this is a large factor $v/v_{th}$ bigger than the typical wind 
expansion rate $dv/dr \approx v/R_{\ast}$, a small perturbation 
at the wind base would, within this lineary theory,
be amplified by an enormous factor, of order 
$e^{v/v_{th}} \approx e^{100}$!

\begin{figure}
\includegraphics[width=11.5cm]{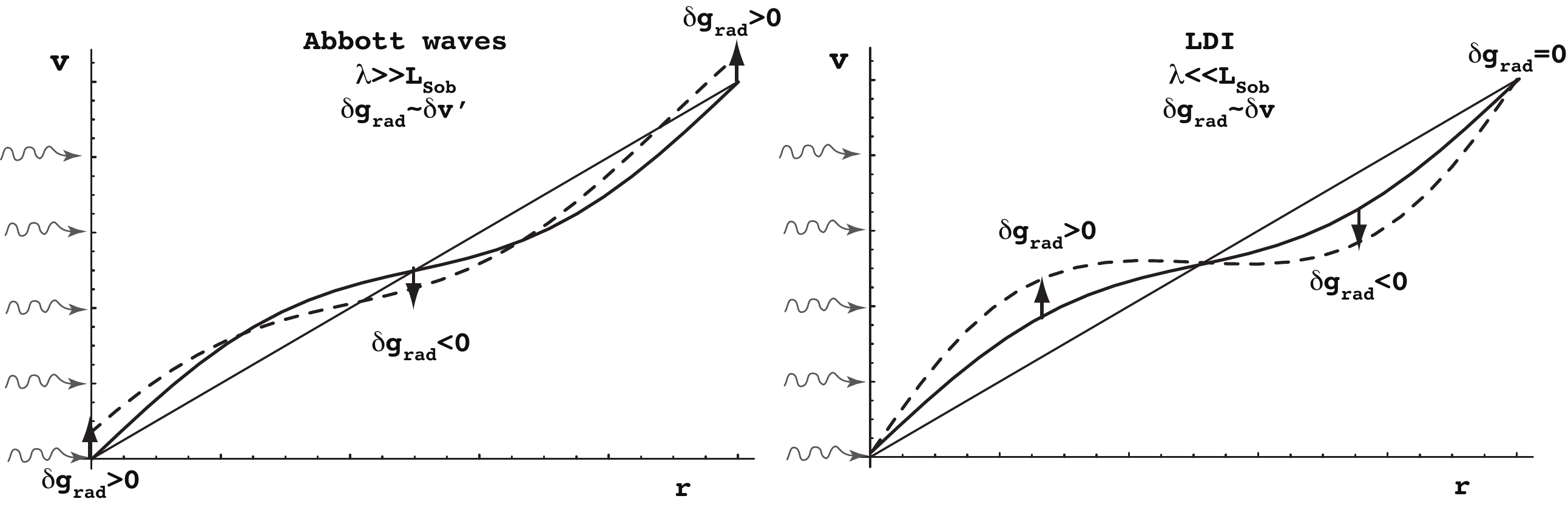}
\caption{
Illustration of the scaling for perturbed line-acceleration $\delta g_{rad}$ with velocity perturbation $\delta v$, showing how this goes from being proportional to the perturbed velocity gradient $\delta v' \sim ik \delta v$ in the long-wavelength Sobolev limit appropriate for Abbott waves, to scaling directly with the perturbed velocity $\delta v$ in the short-wavelength limit of the Line-Deshadowing Instability (LDI).
}
\label{fig:abbldi}
\end{figure}

\subsubsection{Numerical Simulations of Instability-Generated Wind Structure}
\label{sec:ldisims}
 \label{sec5.2.2}

Numerical simulations of the nonlinear evolution require
a non-Sobolev line-force computation on a spatial grid that spans the 
full wind expansion over several $R_{\ast}$, yet resolves
the unstable structure at small scales near and below the Sobolev length.
The first tractable approach 
\citep{OCR88} 
focussed
on the {\em absorption} of the {\em direct} radiation from the stellar core,
accounting now for the attenuation from intervening material by
carrying out a {\em nonlocal integral} for the frequency-dependent
radial optical depth.
Simulations show that because of inward nature of wave propagation
implies an anti-correlation between velocity and density variation,
the nonlinear growth leads to high-speed rarefactions that steepen 
into strong {\em reverse} shocks and compress material into dense 
clumps (or shells in these 1D models) 
 \citep{OCR88}.

The assumption of pure-absorption was criticized by 
 \citet{Lucy84}, 
who pointed out that the interaction of a velocity perturbation with
the background, {\em diffuse} radiation from line-scattering 
results in a {\em line-drag} effect that reduces, and potentially
could even eliminate, the instability associated with the 
direct radiation from the underlying star.
The basic effect is illustrated in figure \ref{fig2}.
The fore-aft ($\pm$) symmetry of the diffuse
radiation leads to cancellation of the  $g_{+}$ and $g_{-}$
force components from the forward and backward streams, as
computed from a  line-profile with frequency centered on the local
comoving mean flow.
Panel b shows that the Doppler shift associated with the
velocity perturbation $\delta v$ breaks this symmetry, and leads to
stronger forces from the component opposing the perturbation.
 
Full linear stability analyses accounting for scattering effects 
\citep{OR85}
show the fraction of the direct instability that is canceled by the
line-drag  of the perturbed diffuse force depends on the ratio of the 
scattering source function $S$ to core intensity $I_{c}$,
\begin{equation}
s = \frac{r^{2}}{R_{\ast}^{2}} \, \frac{2S}{I_{c}} 
\approx \frac{1}{1 + \mu_{\ast}}
~~~ ; ~~~   \mu_{\ast} \equiv \sqrt{1-R_{\ast}^{2}/r^{2}}
\, ,
\label{sdef}
\end{equation}
where the latter approximation applies for the optically thin form
$2S/I_{c} = 1-\mu_{\ast}$.
The net instability growth rate thus becomes
\begin{equation}
\Omega (r) \approx \frac{g_{cak}}{v_{th}}  
\frac{ \mu_{\ast} (r) }{ 1+ \mu_{\ast} (r) }
\, .
\label{omnet}
\end{equation}
This vanishes near the stellar surface, where $\mu_{\ast}= 0$,
but it approaches half the pure-absorption rate far from the star, 
where $\mu_{\ast} \rightarrow 1$.
This implies that the outer wind is still very unstable, with
cumulative growth of ca.\ $v_{\infty}/2v_{th} \approx 50$ e-folds.

Most efforts to account for scattering line-drag in simulations of the
nonlinear evolution of the instability have centered on a 
{\em Smooth Source Function} (SSF) approach 
\citep{Owocki91, Feldmeier95, OP96, OP99}.
This assumes that averaging over frequency and angle makes the
scattering source function relatively insensitive to flow
structure, implying it can be pulled out of the integral in the formal
solution for the diffuse intensity.
Within a simple {\em two-stream} treatment of the line-transport,
the net diffuse line-force then depends on the {\em difference} in the 
{\em nonlocal} escape probabilities $b_{\pm}$ associated with 
forward (+) vs.\ backward (-) {\em integrals} of the frequency-dependent
line-optical-depth.

In the Sobolev approximation, both the forward and backward integrals 
give the same form, viz.\ 
$b_+ \approx b_-$,
leading to the net cancellation of the Sobolev diffuse force.
But for perturbations on a spatial scale near and below the 
Sobolev length, the perturbed velocity breaks the forward/back
symmetry (figure \ref{fig2}b), 
leading to perturbed diffuse force that
now scales in proportion to the {\em negative} of the perturbed velocity,
and thus giving the diffuse line-drag that reduces the
net instability by the factors given in (\ref{sdef}) and (\ref{omnet}).


\begin{figure}
\includegraphics[width=11.75cm]{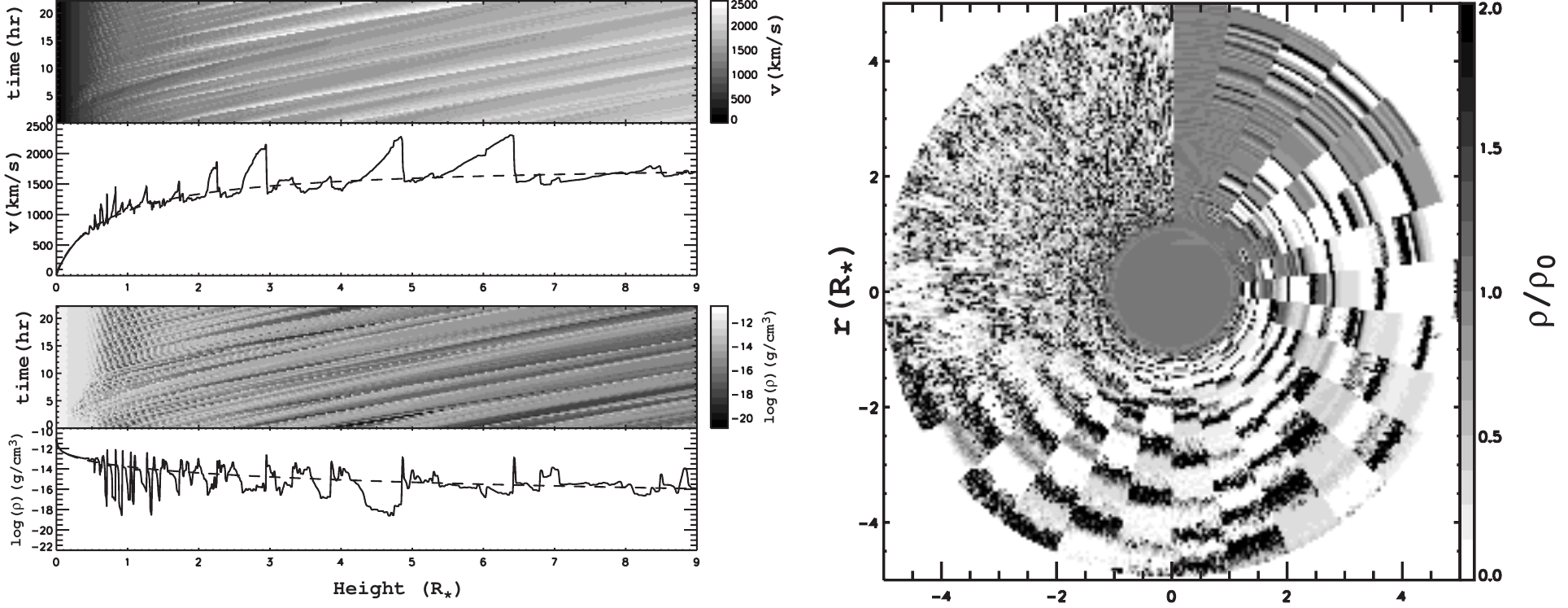}
\caption{
Left: Results of 1D Smooth-Source-Function (SSF) simulation of the 
line-deshadowing instability.
The line plots show the spatial variation of 
velocity (upper) and density (lower) at a fixed, arbitrary
time snapshot.
The corresponding grey scales show both the time (vertical axis)
and height (horizontal axis) evolution.
The dashed curve shows the corresponding smooth, steady CAK model.
Right: For 2DH+1DR SSF simulation, grayscale representation for the
density variations 
rendered as a time  sequence of 2-D wedges of the simulation model azimuthal 
range  $\Delta \phi = 12^{\rm o}$ stacked 
clockwise from the vertical in intervals of 4000~sec from the CAK
initial condition.
}
\label{fig3}
\end{figure}

The left panel of figure \ref{fig3} illustrates the results of a 1D SSF 
simulation, starting from an initial condition set by smooth, steady-state 
CAK/Sobolev model (dashed curves).
Because of the line-drag stabilization of the driving near the star 
(eqn. \ref{omnet}), the wind base remains smooth and steady.
But away from the stellar surface, the net strong instability leads to extensive 
structure in both velocity and density, roughly straddling the CAK steady-state.
Because of the backstreaming component of the diffuse line-force
causes any outer wind structure to induce small-amplitude fluctuations
near the wind base, the wind structure, once initiated, is  ``self-excited'', 
arising spontaneously without any explict perturbation from the
stellar boundary.

In the outer wind, the velocity variations become highly nonlinear and 
nonmonotonic, with amplitudes approaching $1000$~km/s, leading to 
formation of strong shocks. 
However, these high-velocity 
regions have very low density, and thus represent only very little material.
As noted for the pure-absorption models,
this anti-correlation between velocity and density arises because the 
unstable linear waves that lead to the structure have an {\it inward} 
propagation relative to the mean flow.
For most of the wind mass, the dominant overall effect of the instability 
is to concentrate material into dense clumps.
As discussed below, this 
can lead to
overestimates in the mass loss rate from
diagnostics 
that scale with the square of the density.


\begin{figure}
    \includegraphics[width=11.75cm]{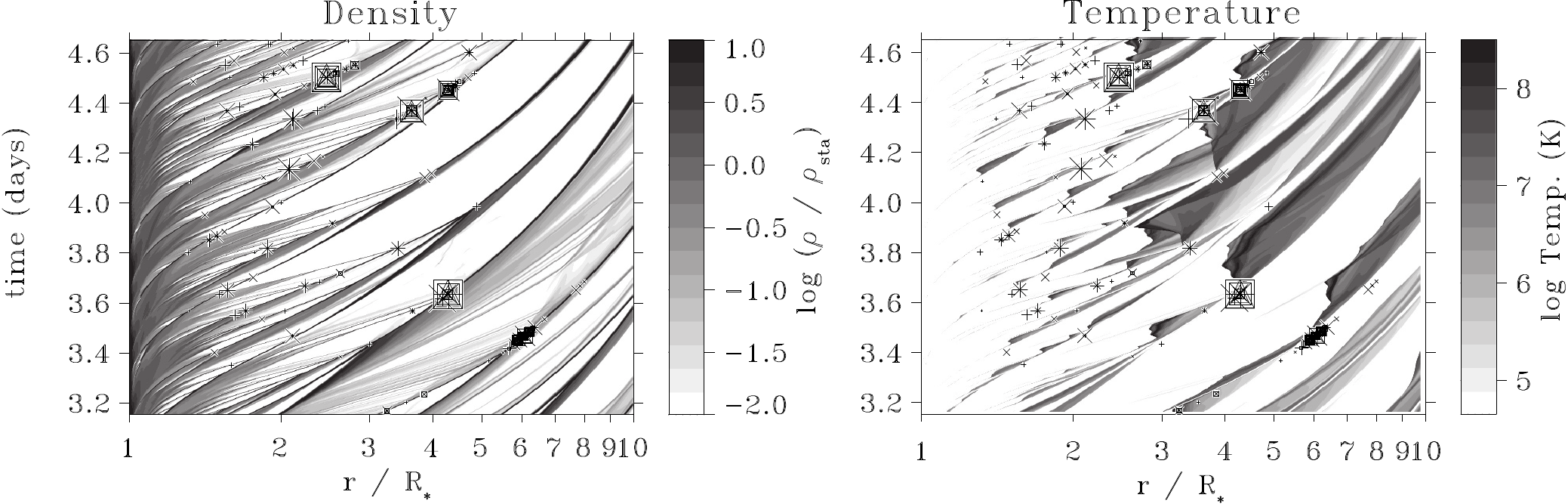}
\caption{
Greyscale rendition of the evolution of wind density and temperature,
for time-dependent wind-instability models with structure formation triggered 
by photospheric perturbations. 
The boxed crosses identify localized region of clump-clump collision that 
lead to the hot, dense gas needed for a substantial level of soft 
X-rays emission.
}
\label{fig4}
\end{figure}

The presence of multiple, embedded strong shocks suggests a potential 
source for the soft X-ray emission observed from massive star winds;
but the rarefied nature of the high-speed gas implies that this
self-excited structure actually
feeds very little material through 
the strong shocks needed to heat gas to X-ray emitting temperatures.
To increase the level of X-ray emission,
\citet{Feldmeier97},
introduced intrinsic perturbations at the wind base, assuming 
the underlying stellar photosphere has a turbulent 
spectrum of compressible sound waves characterized by abrupt phase 
shifts in velocity and density.
These abrupt shifts seed wind variations that, when amplified by the 
line-deshadowing instabilty, now include substantial velocity 
variations among the dense clumps.
As illustrated in figure \ref{fig4}, when these dense clumps 
collide, they induce regions of relatively dense, hot gas which 
produce localized bursts of X-ray emission.
Averaged over time, these localized regions can 
collectively yield X-ray emission with a brightness and spectrum that 
is comparable to what is typically observed from such hot stars.
%

Because of the computational expense of carrying out nonlocal optical depth 
integrations at each time step, such SSF instability simulations have 
generally been limited to just 1D.
More realistically, various kinds of thin-shell instabilities 
\citep{Vishniac94, Kee14} 
can be expected to break up the structure 
into a complex, multidimensional form.
A first step to modelling both radial and lateral structure
is to use a restricted  ``2D-H+1D-R'' approach \citep{Dessart03}, extending the 
hydrodynamical model to 2D in radius and azimuth, 
but still keeping the  1D-SSF radial integration for the inward/outward 
optical depth within each azimuthal zone.
The right panel of figure \ref{fig3} shows
the resulting 2D density structure within a
narrow ($12^{\rm o}$) wedge, with the time evolution rendered 
clockwise at fixed time intervals of 4000 sec starting from the CAK
initial condition at the top.
The line-deshadowing instability is first manifest 
as  strong radial velocity variations and associated density compressions 
that initially extend nearly coherently across the full azimuthal range of 
the computational wedge.

But as these initial ``shell'' structures are accelerated outward, 
they become progressively  disrupted by Rayleigh-Taylor or thin-shell 
instabilities  that operate in azimuth down to the grid scale 
$d \phi = 0.2^{\rm o}$.
%
Such a 2DR+1DH approach may well exaggerate the level of variation on small 
lateral scales.
The lack of {\it lateral} integration needed to compute 
an azimuthal component of the diffuse line-force means that the model 
ignores a potentially strong net lateral line-drag that should 
strongly damp azimuthal velocity perturbations on scales below the lateral
Sobolev length $l_{0} \equiv r v_{th}/v_{r}$ 
 \citep{Rybicki90}.
Presuming that this would inhibit development of lateral instability at 
such scales, then any lateral breakup would be limited to a minimum 
lateral angular scale of $\Delta \phi_{min} \approx l_{0}/r = 
v_{th}/v_{r} \approx 0.01 \, {\rm rad} \approx 0.5^{\rm o}$.
Further work is needed to address this issue through explicit 
incorporation of the lateral line-force and the associated line-drag 
effect.

\subsection{Clumping, Porosity and Vorosity: Implications for Mass Loss
Rates}
 \label{sec5.3}

Both the 1D and 2D SSF simulations thus predict a wind with extensive 
structure in both velocity and density.
A key question then is how such structure might affect the various
wind diagnostics that are used to infer the mass loss rate.
Historically such wind clumping  has
been primarily considered for its effect on diagnostics that scale
with the square of the density,
The strength of such diagnostics is enhanced in a clumped wind,
leading to an overestimate of the wind mass loss rate that scales with
$\sqrt{f_{cl}}$, where the clumping factor
$f_{cl} \equiv \left < \rho^{2} \right >/
\left < \rho \right >^{2} $, with angle brackets denoting a local averaging
over many times the clump scale.
For strong density contrast between the clump and interclump medium,
this is just inverse of the clump volume filling factor, 
i.e. $f_{cl} \approx 1/f_{vol}$.
1D SSF simulations by \citet{Runacres02}
generally find $f_{cl}$ increasing 
from unity at the structure onset radius $\sim 1.5 R_{\ast}$, peaking at a value
$f_{cl} \gtrsim 10$ at $r \approx 10 R_{\ast}$, with then a slow  outward decline to $\sim 5$
for $r \sim 100 R_{\ast}$.

These thus imply that thermal IR and radio emission formed in
the outer wind $r \approx 10-100 R_{\ast}$ may overestimate mass loss
rates by a factor 2-3.
The 2D models of \citet{Dessart03, Dessart05}
find a similar variation,
but somewhat lower peak value,  and thus a lower
clumping factor than in 1D models, with a peak value of about 
$f_{cl} \approx 6$, apparently from the reduced collisional compression from 
clumps with different radial speeds now being able to pass by each  other.
But in both 1D and 2D models, the line-drag near the base means that
self-excited, intrinsic structure does not appear till $r \gtrsim 1.5 $, 
implying little or no clumping effect on $H \alpha$ line emission
formed in this region.
It should be stressed, however, that this is not necessarily a very
robust result, since turbulent perturbations at the wind base, and/or a
modestly reduced diffuse line-drag, might lead to onset of clumping
much closer to the wind base.


If clumps remain optically thin, then they have no effect on
single-density diagnostics, like the bound-free absorption of X-rays.
The recent analysis by Cohen et al. (2010) of the X-ray line-profiles 
observed by Chandra from $\zeta$-Pup indicates matching the relatively
modest skewing of the profile requires mass loss reduction of about a 
factor 3 from typical density-squared diagnostic value.
However, 
a key issue here is whether the individual clumps might
become {\em optically thick} to X-ray absorption.
In this case, the self-shadowing of material within the clump can lead 
to an overall reduction in the effective opacity of the clumped medium
\citep{OGS04, Oskinova07},
\begin{equation}
\kappa_{eff} = \kappa \frac{1 - e^{-\tau_{cl}}}{\tau_{cl}}
\, ,
\label{kapeffdef}
\end{equation}
where $\kappa$ is the microscopic opacity, and the
optical thickness for clumps of size $\ell$ is 
$\tau_{cl} = \kappa \rho \ell f_{cl}$.
The product $\ell f_{cl} \equiv h $ is known as the {\em porosity
length}, which also represents the {\em mean-free-path} between clumps.
A medium with optically thick clumps is thus porous, with an opacity reduction factor
$\kappa_{eff}/\kappa = 1/\tau_{cl} = 1/\kappa \rho h$.

However, it is important to emphasize that getting a significant
porosity decrease in the {\em continuum} absorption of a wind can be quite
difficult, since clumps must become optically thick near the radius of the
smoothed-wind photosphere, implying a collection of a
substantial volume of material into each clump, and so
a porosity length on order the local radius.
\citet{OC06}
showed in fact that a substantial porosity reduction 
the absorption-induced asymmetry of X-ray line profiles required such 
large porosity lengths $h \sim r$.
Since the LDI operates on perturbations at the scale of the Sobolev
length $l_{sob} \equiv v_{th}/(dv/dr) \approx ({v_{th}/v_{\infty}})
R_{\ast} \approx R_{\ast}/300$,
the resulting structure is likewise very small scale,
as illustrated in the 2D SSF simulations in figure~\ref{fig3}.
Given the modest clumping factor $f_{cl} \lesssim 10$, it seems clear
that the porosity length is quite small, $h < 0.1 r$, 
and thus that porosity from LDI structure is not likely to be an important 
factor
for continuum processes like bound-free absorption of X-rays.

The situation is however quite different for {\em line} absorption,
which can readily be  optically thick in even a smooth wind, with
{\em Sobolev optical depth} 
$\tau_{sob} = \kappa_{l} \rho v_{th}/(dv/dr) = \kappa_{l} \rho
l_{sob} > 1$.
In a simple model with a smooth velocity law but material collected
into clumps with volume filling factor $f_{vol}=1/f_{cl}$, this clump 
optical depth would be even larger by a factor $f_{cl}$.
As noted by 
\citet{Oskinova07},
the escape of radiation in the gaps between the thick clumps might then
substantially reduce the effective line strength, and so 
help explain the unexpected weakness of PV lines observed by FUSE 
\citep{Fullerton06},
which otherwise might require a substantial, factor-ten or more reduction 
in wind mass loss rate.

\begin{figure*}[!t]
\begin{center}
\includegraphics[width=1.0\textwidth]{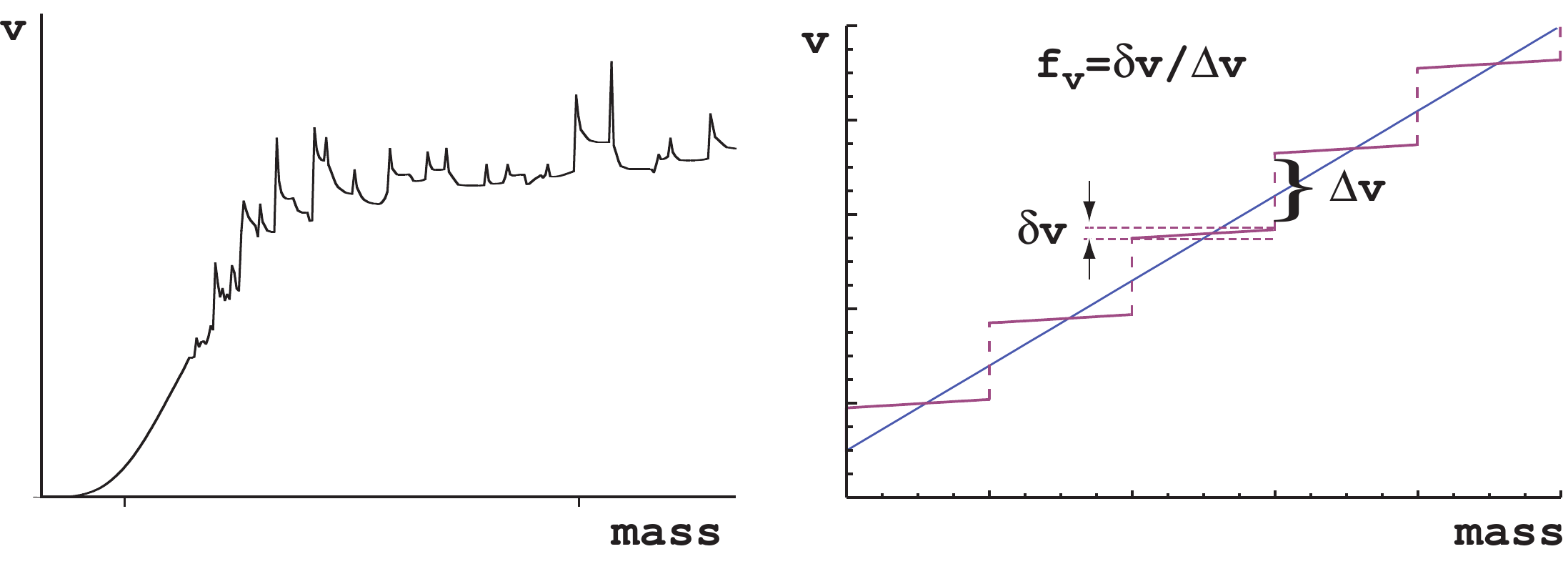}
\caption{
Left: Self-excited velocity structure arising in a 1D SSF simulation of the
line-driven instability, plotted versus a mass
coordinate, $M(r) = \int_{R}^{r} 4 \pi \rho r'^{2} \, dr' $.
Note the formation of velocity plateaus in the outer regions of the
wind.
Right: Velocity vs.\ mass in a wind seqment with structure described by
a simplified velocity staircase model with multiple large steps
$\Delta v$ between plateaus of width $\delta v$.
Here the associated velocity clumping factor 
$f_{vel} \equiv \delta v/\Delta v = 1/10 $.
The straight line represents the corresponding smooth 
CAK/Sobolev model.
}
\label{owocki-fig5}
\end{center}
\end{figure*}


But instead of {\em spatial} porosity, the effect on lines is better
characterized as a  kind of velocity porosity , or ``{\em vorosity}'',
which is now relatively insensitive to the spatial scale of wind structure
\citep{Owocki08}.
%
%
The left panel of figure \ref{owocki-fig5} illustrates the typical
result of 1D dynamical simulation of the wind instability, 
plotted here as a time-snapshot of velocity vs.\ a {\em mass} coordinate,
instead of radius.
The intrinsic instability of line-driving leads to a 
substantial velocity structure, with narrow peaks corresponding to spatially
extended, but tenuous regions of high-speed flow; these bracket dense, 
spatially narrow clumps/shells that appear here as nearly flat, extended
velocity plateaus in mass.
The right panel of figure \ref{owocki-fig5} illustrates a simplified, 
heuristic model of such wind structure for a representative wind section,
with the velocity clumping now represented by a simple ``staircase'' structure,
compressing the wind mass into discrete sections of the wind velocity law, 
while evacuating the regions in between;
the structure is characterized by a
``velocity clumping factor'' $f_{vel}$, set by the ratio between the 
internal velocity width $\delta v$ to the 
velocity separation $\Delta v$ of the clumps.
The straight line through the steps represents the
corresponding smooth wind flow.


The effect of the velocity structure on the line-absorption  profile 
depends on the local Sobolev optical depth, 
which scales with the inverse of the mass derivative of velocity,
$\tau_{y} \sim 1/(dv/dm)$,
evaluated at a resonance location 
$r_{s}$, where  the velocity-scaled, observer-frame wavelength
$y =-v(r_{s})/v_{\infty}$.
In a smooth wind with Sobolev optical depth $\tau_{y}$, 
the absorption profile is given simply by 
$
A_{y} = 1 - e^{-\tau_{y}} 
$
\citep{Owocki08}.
In the structured model,
the optical thickness of individual clumps
is increased by the inverse of the clumping factor $1/f_{vel}$,
but they now only cover a fraction $f_{vel}$ of the
velocity/wavelength interval.
The net effect on the averaged line profile is to 
{\em reduce} the net aborption by a factor \citep{Owocki08},
\begin{equation}
R_{A} (\tau_{y},f_{vel}) 
= f_{vel} ~ 
\frac{  1 - e^{-\tau_{y}/f_{vel}} }
      { 1 - e^{-\tau_{y}        } }
\, .
\label{owocki-radef}
\end{equation}
Note that for optically thick lines, $\tau_{y} \gg 1$, 
the reduction approaches a fixed value, given in fact by the clumping factor,
$R_{A} \approx f_{vel}$.
If the smooth-wind line is optically thin, $\tau_{y} \ll 1 $, then 
$R_{A} (\tau_{y}, f_{vel}) \approx (1 - e^{-\tau_{y}/f} )/
(\tau_{y}/f_{vel})$, which is quite analogous to the opacity reduction for
{\em continuum} porosity (eqn.\ \ref{kapeffdef}),
if we just substitute for the clump optical depth, 
$\tau_{c} \rightarrow \tau_{y}/f_{vel}$.

But a key point here is that, unlike for the continuum case, 
the {\em net reduction in line absorption no longer depends on
the spatial scale} of the clumps.
Instead one might think of this velocity clumping model as a kind 
of velocity form of the standard venetian blind, 
with $f_{vel}$ representing the fractional projected covering factor of the 
blinds relative to their separation.
The $f_{vel}=1$ case represents closed blinds that effectively block the
background light, while small $f_{vel}$ represent cases when the blinds
are broadly open, letting through much more light.

Further discussion of the potentially key role of wind clumping and vorosity for determining wind mass loss rates is given by \citet{Sundqvist11}.

\section{Continuum-Driven Mass Loss from Super-Eddington LBVs}
 \label{sec6}

\subsection{Lack of Self-Regulation for Continuum Driving}
\label{sec:windenergy}
 \label{sec6.1}

Despite the extensive instability-generated structure in the outer regions of line-driven winds, their overall mass loss is quite steady, and can persist throughout the lifetime of even moderately massive stars.
But VMS show occasional episodes of much stronger mass loss, known generally as giant eruption LBVs, commonly characterized by a radiative luminosity that exceeds even the classical electron-scattering  Eddington limit,
and lasting for up to about a decade.
The energy source and trigger of such eruptive LBVs is uncertain, and could even have an explosive character seated in the deep stellar interior; but their persistence for much longer than the dynamical time scale of few days suggests they can be at least partly modeled as a quasi-steady stellar wind, though now driven by {\em continuum} opacity through electron scattering of their super-Eddington luminosity.

A key issue for such continuum-driven wind models is that they lack a natural self-regulation. In line-driven winds,  the self-absorption and saturation of line flux defers the onset of line-acceleration to a relatively low-density near-surface layer, thus limiting  the associated mass flux to a value that can be sustained to full escape from star by the expansion-desaturated line-driving in the outer wind.  For continuum driving by a gray opacity like electron scattering, the bolometric flux does not saturate, keeping the radiative force strong even in dense, optically thick layers well below the photospheric surface.
As discussed below, a mass outflow initiated from such deep, dense layers becomes difficult to sustain with the finite   energy flux available from the stellar interior, and this can lead to flow stagnation and infall, with extensive variability and spatial structure.

\subsection{Convective Instability of a Super-Eddington Interior}
 \label{sec:conv}
  \label{sec6.2}

It should be emphasized, however, that locally exceeding the Eddington limit
need {\it not} necessarily lead to initiation of a mass outflow.
As first shown by \citet{Joss73},
in a stellar envelope 
allowing the Eddington parameter $\Gamma \rightarrow 1$ 
generally implies through the Schwarzschild criterion that material 
becomes {\it convectively unstable}.
Since convection in such deep layers is highly efficient at
transporting the energy, the contribution from the radiative flux is reduced, 
thereby lowering the associated radiative Eddington
parameter away from unity.

This suggests that, even in a star that formally exceeds the Eddington limit, a radiatively driven outflow could only be initiated
{\em outside} the region where convection is {\em efficient}.
An upper bound to the convective energy flux is set by
\begin{equation}
F_{conv} \approx v_{conv} \, l \, dU/dr \lesssim a \, H \, 
dP/dr \approx a^3 \rho , 
\end{equation} 
where $v_{conv}$, $l$, and $U$ are the convective velocity, mixing length,
and internal energy density, and $a$, $H$, $P$, and $\rho$ are the sound speed,
pressure scale height, pressure, and mass density.
Setting this maximum convective flux equal to the total stellar energy 
flux $L_{}/4 \pi r^2$ yields an estimate for the maximum mass loss rate 
that can be initiated by radiative driving,
\begin{equation}
\frac{\dot M} \le {L_{} }{ a^2 } \equiv {\dot M}_{max,conv} 
\, .
\label{mdmaxconv}
\end{equation}
This is a very large rate, generally well in excess of the fundamental limit
set by the energy available to lift the material out of the
star's gravitational potential. In terms of the escape speed $\vesc \equiv \sqrt{GM/\Rstar}$ from the stellar surface radius $\Rstar$, this can be written as
\begin{equation}
{\dot M}_{tir} = \frac { L_{} }{ v_{esc}^{2}/2 } 
= \frac{ L_{} }{ GM_{}/\Rstar } 
= 0.032 \, \frac{M_{\odot} }{ {\rm yr}} ~ 
\frac{L_{}}{10^{6} L_{\odot}}
\frac{M_{\odot}/R_{\odot}}{M_{}/R_{}}
\, .
\label{mdtir}
\end{equation}
 where $L_{6} \equiv L_{}/10^{6} L_{\odot}$.
As indicated by the subscript, this is commonly referred to as the {\em photon tiring} limit, since the radiation driving such a mass loss would lose energy, or become ``tired'', from the work done to lift the material against gravity.
Since generally $a \ll \vesc$, 
a mass flux initiated from the radius of inefficient convection would greatly exceed the photon tiring limit,
implying again that any such outflow would necessarily stagnate at some 
finite radius. 

\subsection{Flow Stagnation from Photon Tiring}
 \label{sec6.3}
 
To account for the reduction in the the radiative luminosity $L(r)$ due to the net work done in lifting and accelerating the wind from the stellar radius $\Rstar$ to a local  radius $r$ with wind speed $v(r)$, we can write
\beq
L(r) =  L_o  - 
\Mdot  
\left [ \frac{ v(r)^2 }{ 2 } + \frac{G\Mstar }{ \Rstar } - \frac{G\Mstar }{ r } \right ] 
\, ,
\eeq
where $L_o \equiv L(\Rstar) $ is the radiative luminosity at the wind base.
For the dimensionless equation of motion (\ref{dimlesseom}),
\beq
\left ( 1 - \frac{\ws}{w} \right )
\, w' = \Gamma (x)- 1 \, ,
\nonumber
\label{dimlesseom2}
\eeq
the associated Eddington parameter depends on the scaled wind energy $w$ and inverse radius coordinate $x$,
\beq
\Gamma(x) =  \Gamma_o (x) [ 1 - m (w+x) ]  \, ,
\label{eq:gamtir}
\eeq
where $\Gamma_o (x) \equiv \kappa(x) L_o/4 \pi GMc$.
Here the gravitational ``tiring number'',
\beq
m \equiv \frac{\Mdot }{ \Mdot_{tir}}
= \frac{ \Mdot G\Mstar }{ \Lstar \Rstar } 
  \approx 0.012 \, \frac{ \Mdot_{-4} V_{1000}^2 }{ L_6 } 
\, ,
\label{mtirdef}
\eeq
characterizes the {\em fraction} of radiative energy lost in lifting
the wind out of the stellar gravitational potential.
The last expression allows easy evaluation of the likely importance of 
photon tiring for characteristic scalings, 
where $\Mdot_{-4} \equiv \Mdot/10^{-4}\, \Msun/yr$, 
$L_6 \equiv \Lstar/10^6 \,\Lsun$, 
and $V_{1000} \equiv v_{esc}/1000\, km/s $
$\approx 0.62 \,(\Mstar/\Rstar)/(\Msun/\Rsun)$.
In particular, for the typical CAK wind scalings of line-driven winds, $m < 0.01$, justifying the neglect of photon tiring in the CAK model discussed in \S \ref{sec5}.

More generally for cases with non-negligible tiring numbers $m \ltwig 1$, the equation of motion   (\ref{dimlesseom2}) can be solved using integrating factors, yielding an explicit solution
for $w(x)$ in terms of the integral quantity 
${\bar \Gamma}_o (x) \equiv \int_0^x dx' \Gamma_o (x')$,
\begin{equation}
w(x) = -x + \frac{1 }{ m } \, 
\left [ 1 - e^{-m {\bar \Gamma}_o (x) } \right ] + \ws
\, ,
\label{wsoln}
\end{equation}
where for typical hot-star atmospheres the sonic point boundary value 
is very small, $w(0) = \ws < 10^{-3}$.


\begin{figure}
\includegraphics[width=5.5cm]{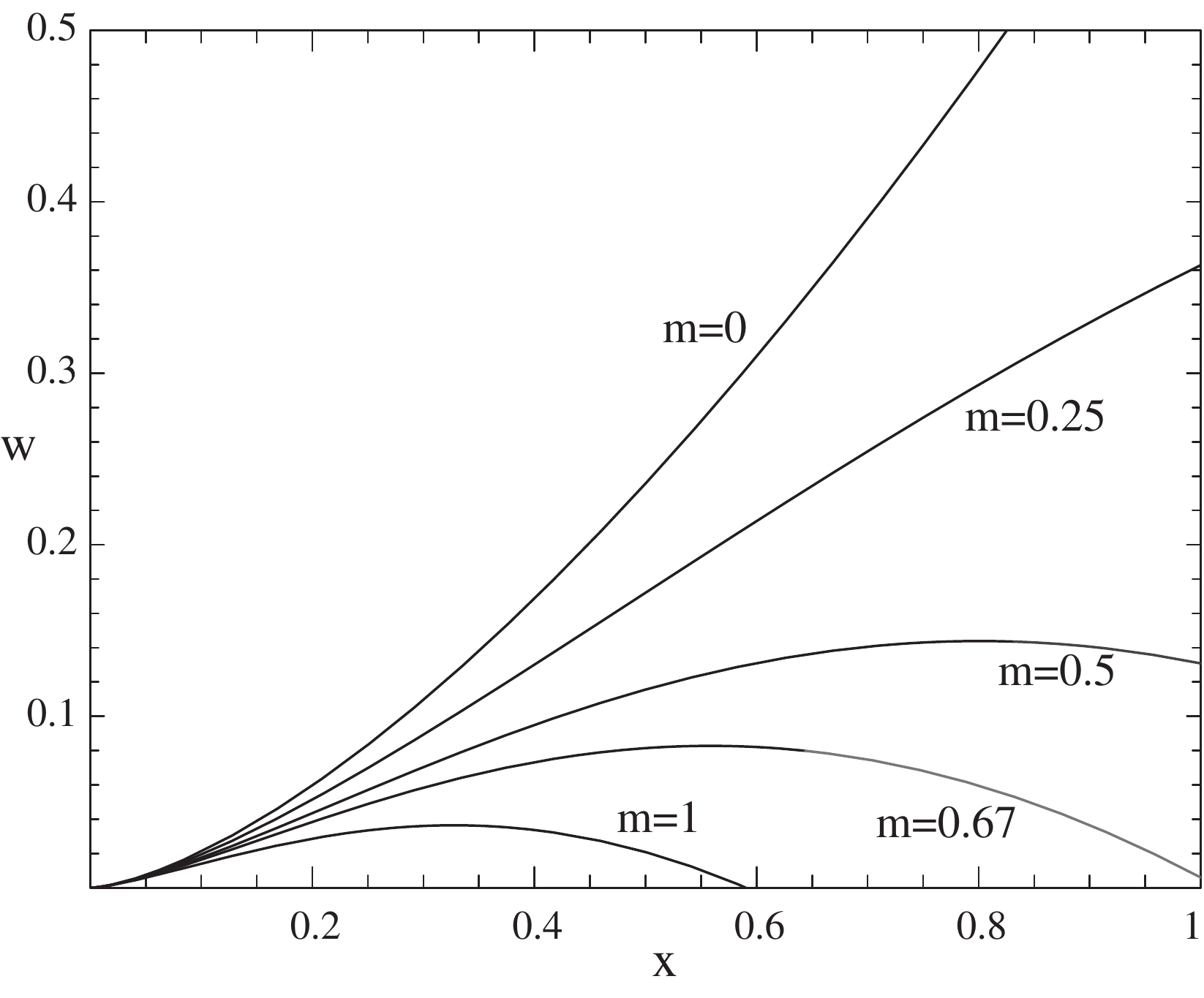}
\includegraphics[width=5.5cm]{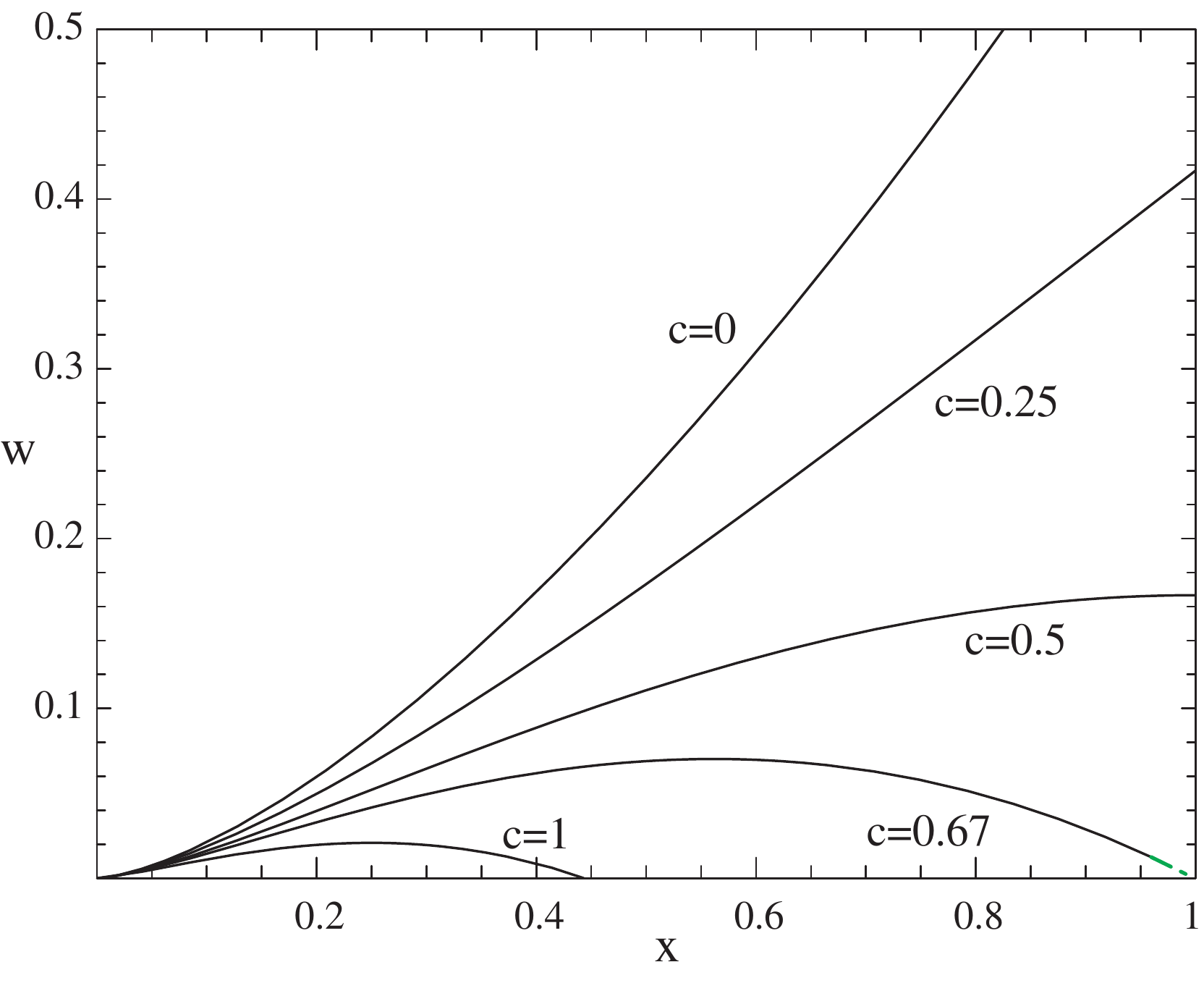}
\caption{
a. Wind energy $w$ vs.~scaled inverse radius $x$($\equiv 1-\Rstar/r$),
plotted for Eddington parameter $\Gamma_o (x) = 1 + \sqrt{x}$
with various photon tiring numbers $m$.
b. Same as (a), except for weak tiring limit $m \ll 1$, and for 
various constants $c$ in the Eddington parameter scaling
$\Gamma_o (x)=1+\sqrt{x}-2c x$.
}
\label{fig:tiring}
\end{figure}

As a simple example, consider the case\footnote{The choice of these 
functions is arbitrary, to illustrate the photon-tiring effect within 
a simple model. More physically motivated models based on a medium's  
porosity are presented in \S \ref{sec6.5}} 
with $\Gamma_{0} (x) = 1 + \sqrt{x}$, for which ${\bar \Gamma_{c}} = x + 2x^{3/2}/3$.
Fig.~\ref{fig:tiring}a plots solutions $w(x)$ vs.~$x$ from eqn.~(\ref{wsoln})
with various $m$.
For low $m$, the flow reaches a finite speed at large radii ($x=1$), 
but for high $m$, it curves back, stopping at some finite {\it stagnation}
point $x_s$, where $w(x_s) \equiv 0$.  
The latter solutions represent flows for which the mass
loss rate is too high for the given stellar luminosity to be able 
to lift the material to full escape at large radii.
In a time-dependent model, such material can be expected to accumulate
at this stagnation radius, and eventually fall back to the star (see \S \ref{sec:conwinds} and figure \ref{fig:por-tir}).

Figure \ref{fig:tiring}b shows that, even without photon tiring, a similar stagnation can occur from an 
outward decline in $\Gamma$ after an initially super-Eddington driving, as might occur, for example, in the region above the iron bump.
%
Again, a general point here is that even if a super-Eddington condition $\Gamma > 1$ initiates an outflow, it does not guarantee that material will escape to large radii in a steady-state wind.

\subsection{Porosification of VMS Atmospheres by Stagnation and Instabilities}
\label{sec:rad_insta}
 \label{sec6.4}

The stagnation implied by the above simple steady 1D models can lead to a complex, time-dependent, 
3D pattern of outflows and inflows in actual VMS atmospheres.
Moreover, dating back to early work by \citet{Spiegel76, Spiegel77}, there have been 
speculations that atmospheres supported by radiation pressure would 
likely exhibit instabilities not unlike that of Rayleigh-Taylor, 
associated with the support of a heavy fluid by a lighter one, 
leading to formation of  ``photon bubbles''. 
Quantitative stability analyses \citep{Spiegel99,Shaviv01} indicate that even a simple case of a pure ``Thomson
atmosphere'' ---i.e., supported by Thomson scattering of radiation by
free electrons--- could be subject to intrinsic instabilities for
development of lateral inhomogeneities, with many
similar properties to the excitation of strange mode pulsations
\citep{Glatzel94,Pap97} discussed in \S \ref{sec4.2}.
If magnetic fields are introduced, even more instabilities come 
into play \citep{Arons92, Gammie98, Begelman02, Blaes03}.  

The general upshot is that the atmospheres of VMS should have extensive variability and spatial structure, characterized by strong density inhomogeneities over a wide range in length scales.
In deeper layers with a high mean density, we can expect that that many of the largest and/or densest clumps should individually become optically thick, forcing the radiation flux to preferentially diffuse through relatively low-density channels {\em between} the clumps.

This is the same spatial ``porosity'' effect discussed in \S \ref{sec5.3}, which {\em reduces} the effective coupling of the gas and radiation in the deeper layers. In the present context the net result is to keep these inner dense layers gravitationally bound even when the radiative flux exceeds the Eddington limit.
This defers the onset of a continuum driven wind outflow to a higher, lower-density layer where the clumps are becoming optically thin, resulting in a more moderate mass loss rate that can be more readily sustained.

Figure\ \ref{fig:cartoon-structure} illustrates the expected overall  structure of stars with a super-Eddington luminosity, wherein porosity-regulated continuum opacity drives a quasi-steady wind  from a stably bound atmospheric base. The distinct physical layers from interior to wind are as follows:

\vskip 2mm \noindent 
 {\bf (A)} As elaborated upon in \S\ref{sec:conv}, deep inside the
 star where the density is sufficiently high, any excess flux above
 the Eddington luminosity is necessarily advected through convection.
 Thus, we have a bound layer with $L_{{\rm rad}} < L_{{\rm Edd}} <
 L_{{\rm tot}}$.

\vskip 2mm \noindent 
{\bf (B)} At lower densities, where convection is inefficient,
radiative instabilities necessarily force the atmosphere to
become inhomogeneous.  This reduces the effective opacity and thus
increases the effective Eddington luminosity $L_{\rm eff}$.  In other
words, this layer is bound, not because the flux is lowered (as it
is in the convective regions), but because the opacity is reduced.
Thus here we find $L_{{\rm Edd}} < L_{{\rm rad}} = L < L_{{\rm eff}}$.
 
\vskip 2mm \noindent 
{\bf (C)} Opacity reduction can operate only as long as the
individual clumps are optically thick.  In high layers with lower density,
the clumps lose their opaqueness and so the effective opacity
recovers the microscopic value (and thus $L_{\rm eff}$ to $L_{\rm Edd}$).
A sonic/critical point of a wind will therefore be located where $L =
L_{{\rm eff}} \gtrsim L_{{\rm Edd}}$.  

\vskip 2mm \noindent 
 {\bf (D)} Since the mass loss rate is large, the inner wind is optically
 thick and the radiative photosphere resides in the wind itself, at some radius where
 geometrical dilution eventually makes it become transparent.

\begin{figure}[t]
\begin{center}
\includegraphics[width=3.in]{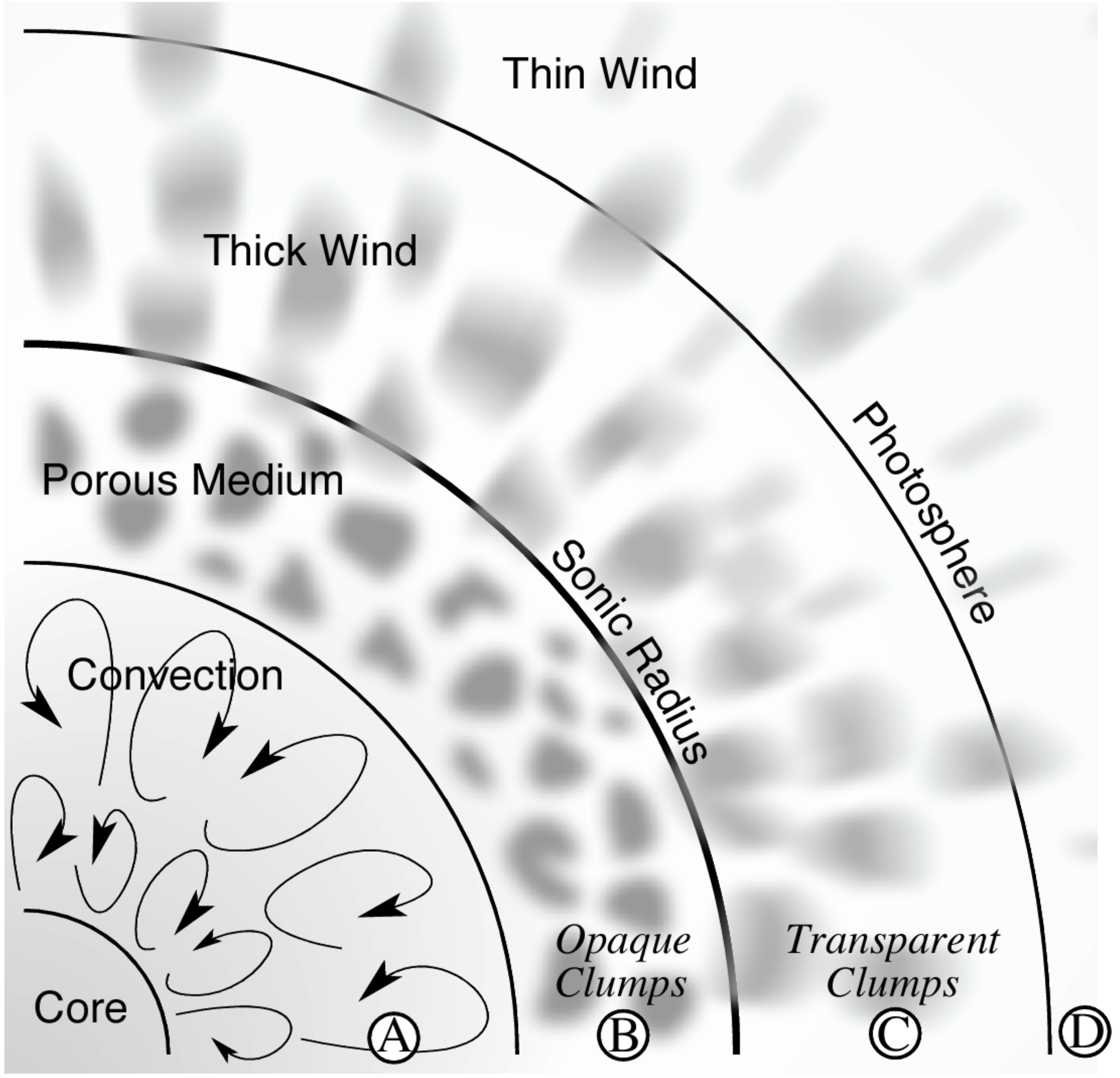} 
 \end{center}
\caption{ The structure of a super-Eddington star. 
The labelled regions are described in the text .  }
\label{fig:cartoon-structure}
\end{figure}

\subsection{Continuum-driven Winds Regulated by Porous Opacity}
\label{sec:conwinds}
 \label{sec6.5}
 
To develop quantitative scalings for such porosity-regulated mass loss, let us again first 
consider a medium in which
material has coagulated into discrete clumps of individual optical
thickness $\tau_{cl} = \kappa \rho_{cl} \ell$, where $\ell$ is the clump scale,
and the clump density is enhanced compared to the mean density of the
medium by a volume filling factor $f = \rho_{cl} / \rho$.
The effective overall opacity of this medium can then be approximated 
by the form given in (\ref{kapeffdef}),
\begin{equation}
   \kappa_\mathrm{eff} \approx \kappa \, \frac{ 1 - e^{-\tau_{cl}}  }{ \tau_{cl}} .
\nonumber
\end{equation}
Note again that in the limit of optically thin clumps ($\tau_{cl} \ll 1$) this 
reproduces the usual microscopic opacity  ($\kappa_\mathrm{eff} \approx \kappa$);
but in the optically thick limit ($\tau_{cl} \gg 1$), the effective 
opacity is reduced by a factor of $1/\tau_{cl}$, thus yielding a medium 
with opacity characterized instead by the clump cross section divided by the 
clump mass ($\kappa_\mathrm{eff} = \kappa/\tau_{cl} = \ell^{2}/m_{cl}$).
The critical mean density at which the clumps become optically thin is 
given by $\rho_{o} = 1/\kappa h$, where $h \equiv \ell/f$ is a
characteristic ``porosity length'' parameter.
A key upshot of this is that the radiative acceleration in such a 
gray, but spatially porous medium would likewise be reduced by a 
factor that depends on the mean density.

More realistically, it seems likely that structure should occur 
with a range of compression strengths and length scales.
Drawing upon an analogy 
with the power-law distribution of line-opacity in the  standard CAK model 
of line-driven winds,
let us thereby consider a {\it power-law-porosity} model in which the 
associated structure has a broad range of porosity length $h$. 
As detailed in 
\citet{OGS04}, 
this leads to an effective Eddington parameter that scales as
\begin{equation}
\Gamma_\mathrm{eff} \approx \Gamma \left ( \frac{ \rho_{o} }{ \rho } \right 
)^{\alpha_{p}} ~~ ; ~~ \rho > \rho_{o}
\, ,
\end{equation}
where $\alpha_{p}$ is the porosity power index (analogous to the CAK 
line-distribution power index $\alpha$),
and 
$\rho_{o} \equiv 1/\kappa h_{o}$, with $h_{o}$ now the 
porosity-length associated with the {\em strongest} 
(i.e. most optically thick) clump.
As discussed by \citet{Sundqvist12b}, a power index $\alpha_{p} =2$ gives 
the same transport as
a simple two-component (clump + void) medium described by Markovian statisitics \citep{Levermore86, Pomraning91}.

In rough analogy with the ``mixing length'' formalism of stellar 
convection, let us assume the basal porosity length $h_{o}$ scales with
gravitational scale height $H \equiv a^{2}/g$. 
Then the requirement that $\Gamma_\mathrm{eff}=1$ at the wind sonic point 
yields a scaling for the mass loss rate scaling with luminosity.
For a canonical case $\alpha_p=1/2$ \citep{OGS04}, we find
\begin{eqnarray}
{\dot M}_{por} (\alpha_p = 1/2)
&=& 4 (\Gamma - 1) \, \frac{L }{ a c} \, \frac{H }{ h_{o}}
\label{Mdporab}
\\
&=& 0.004 (\Gamma-1) \, \frac{M_{\odot} }{ {\rm yr}} \, \frac{L_{6} }{ a_{20} }
\, \frac{H }{ h_{o}}
\, .
\label{Mdporc}
\end{eqnarray}
The second equality gives numerical evaluation in terms of characteristic 
values for the sound speed $a_{20} \equiv a/20$~km/s
and luminosity $L_{6} \equiv L/10^{6} L_{\odot}$.
Comparision with the CAK  scalings (\ref{mdcak}) for a line-driven wind
shows that the mass loss can be substantially higher from a super-Eddington
star with porosity-regulated, continuum driving.
Applying the extreme luminosity 
$L \approx 20 \times 10^{6} \, L_{\odot}$ estimated for the 1840-60
outburst of $\eta$~Carinae,
which implies an Eddington parameter $\Gamma \approx 5$, 
the derived mass loss 
rate for a canonical porosity length of $h_{o} = H$ is 
${\dot M}_{por} \approx 0.32 M_{\odot}$/yr, 
quite comparable to the inferred average $\sim 0.5 M_{\odot}$/yr
during this epoch.

For comparison, a Markov model with $\alpha_p=2$ gives a different, weaker scaling of mass loss with $\Gamma$,
\beq
{\dot M}_{por} (\alpha_p=2)
=  \left ( 1 - \frac{1}{\Gamma} \right ) \, \frac{L }{ a c} \, \frac{H }{ h_{o}} = \frac { {\dot M}_{por} (\alpha_p=1/2)}{4 \Gamma}
\, ,
\label{Mdmarkov}
\eeq
which saturates to a fixed limit for $\Gamma \gg 1$.
To reach the mass loss inferred for $\eta$~Carianae's giant eruption, such a Markov model would need to have a much smaller porosity length, e.g.\ $h_o \approx 0.05 H$.

But overall, it seems that, together with the ability to drive quite fast 
outflow speeds (of order the surface escape speed), the extended porosity 
formalism provides a promising basis for self-consistent dynamical modeling 
of even the most extreme
mass loss outbursts of Luminous Blue Variables, namely those that, 
like the giant 
eruption of $\eta$~Carinae, approach the photon tiring limit.

\subsection{Simulation of Stagnation and Fallback above the Tiring Limit}
 \label{sec6.6}

For porosity models in which the base mass flux {\em exceeds} 
the photon tiring limit,  numerical simulations 
\citet{vanMarle09}
have explored the nature of the resulting complex pattern of infall 
and outflow.
Despite the likely 3-D nature of such flow patterns, 
to keep the computation tractable, this initial exploration assumes 1-D 
spherical symmetry, though now allowing a fully time-dependent density 
and flow speed.
The total rate of work done by the radiation on the
outflow (or vice versa in regions of inflow) is again accounted for by a
radial change of the radiative luminosity with radius,
\begin{equation}
\frac{dL }{ dr} 
= - {\dot m} g_{rad}
= - \kappa_\mathrm{eff} \, \rho v L/c
\, ,
\label{dldr}
\end{equation}
where ${\dot m} \equiv 4 \pi \rho v r^{2}$ is the local mass-flux at
radius $r$, which is no longer a constant, or even monotonically positive,
in such a time-dependent flow.
The latter equality then follows from the definition (\ref{eq:grad}) 
of the radiative acceleration $g_{rad}$ for a gray opacity
$\kappa_\mathrm{eff}$, set here by porosity-modified electron scattering.
At each time step, eq. (\ref{dldr}) is integrated from an assumed
lower boundary luminosity $L(R)$ to give the local radiative 
luminosity $L(r)$ at all radii $r > R$.
Using this to compute the local radiative acceleration,
the time-dependent equations for mass and momentum conservation
are evolved forward to obtain the time and radial variation of 
density $\rho (r,t)$  and flow speed $v (r,t)$. 
(For simplicity, the temperature is fixed
at the stellar effective temperature.)
The base Eddington parameter is $\Gamma=10$,
and the analytic porosity mass flux is 2.3 times the tiring limit.

\begin{figure}[t]
\includegraphics[width=4.7in]{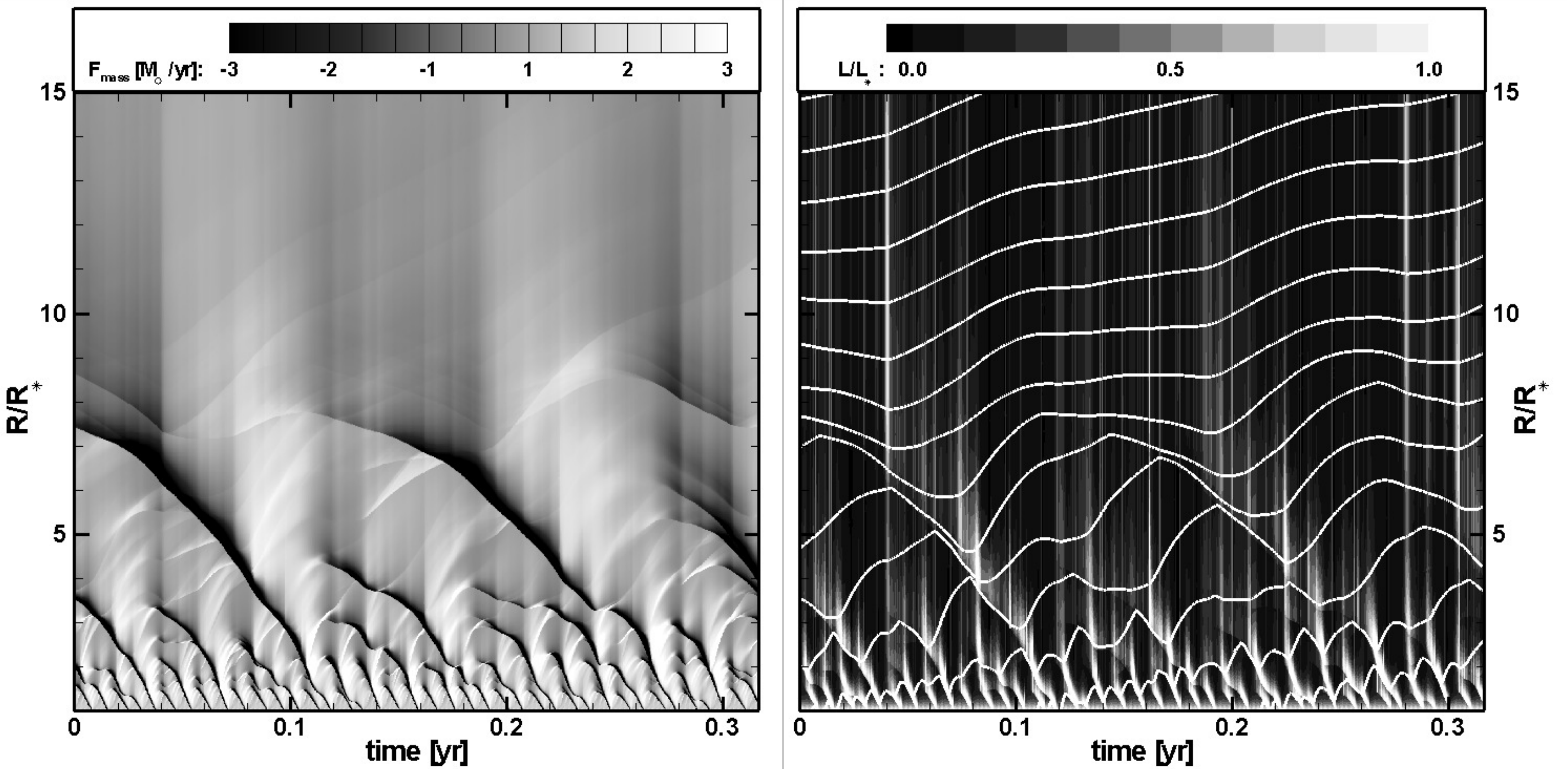} 
\caption{
Grayscale plot of radius and time variation of mass flux (left) and 
luminosity (right) in a time-dependent simulation of a super-Eddington wind 
with a porosity-mediated base mass flux above the photon tiring limit.
The white contours on the right trace the height progression of fixed 
mass shells.
}
\label{fig:por-tir}
\end{figure}

Figure \ref{fig:por-tir} illustrates the flow structure as a function of radius 
(for $r=1$-$15~R$) and time 
(over an arbitrary interval long after the initial condition)
The left panel grayscale shows the local mass flux, in $M_{\odot}/$~yr,
with dark shades representing inflow, and light shades outflow.
In the right panel, the shading represents the local luminosity in
units of the base value, $L(r)/L(R)$, ranging from zero (black) to one 
(white); in addition, the superposed lines represent the radius and
time variation of selected mass shells.

Both panels show the remarkably complex nature of the flow,
with positive mass flux from the base overtaken by a hierarchy of
infall from stagnated flow above. 
However, the re-energization of the radiative luminosity from this
infall makes the region above have an outward impulse.
The shell tracks thus show that, once material reaches a radius 
$r \approx 5 R$, its infall intervals become ever shorter, allowing it
eventually to drift outward.
The overall result is a net, time-averaged mass loss through the
outer boundary that is very close to the photon-tiring limit, 
with however a terminal flow speed  $v_{\infty} \approx 50$~km/s 
that is substantially below the surface escape speed $v_{esc} \approx 600$~km/s.

These initial 1-D simulations thus provide an interesting glimpse into this 
competition below inflow and outflow. 
Of course, the structure in more realistic 2-D and 3-D models is
likely to be even more complex, and may even lead itself to a highly porous 
medium.
But overall, it seems that one robust property of super-Eddington
stars may well be mass loss that is of the order of the photon tiring
limit.

\subsection{LBV Eruptions: Enhanced Winds or Explosions?}
 \label{sec6.7}

The previous section has modeled the eruptive, LBV mass loss of VMS in terms of a
quasi-steady, continuum-driven wind that results from the stellar
luminosity exceeding the Eddington limit.
But an alternative paradigm is that such eruptions might in fact be
point-time ``explosions'' that simply did not have sufficient energy
to completely disrupt the star.

Both paradigms require an unknown energy source, but one important
distinction is that explosions are driven by {\em gas pressure}, 
whereas super-Eddington winds are driven by {\em radiation}.
The two have markedly different timescales.

The overpressure from an explosion propagates through the star on a
very short dynamical time scale, of order $R/a$, where $a$ is the
sound-speed in the very high temperature gas that is heated by the
energy deposition of the explosion.
In supernovae, this sound speed is on the order of the mass ejection 
speed, on the order of 10,000~km/s;
even in a ``failed'' LBV explosion, it would be on the order of the
surface escape speed, 
or a few hundred km/s, implying 
a dynamical time of order the free fall time, 
or just a few hours.
Of course, the release of radiative energy is tied to the
expansion (and later on, radioactive $\beta$-decay), and thus peaks on
a somewhat longer time of a few days or
weeks for supernovae.
But it is difficult to see how such a direct gas-pressure-driven
explosion could be maintained for the years to decade timescale 
inferred for LBV eruptions.

This then is perhaps the key argument for a radiation-driven
model. 
If energy is released in the deep interior, its {\em radiative}
signature can take up to a much longer {\em diffusion} time to reach 
the surface\footnote{Since the luminous stars are likely to be mostly
convective (e.g. \S\ref{sec:conv}), the limiting time scale is that of
the convective diffusion's mixing length time in the stellar cores,
which due to the high density is much longer than the dynamical time
scales.}.
This can be long as a few years.

In contrast to the explosive disruption of supernovae, for LBV eruptions 
the total energy is typically well below the stellar binding energy.
Thus even if this energy were released suddenly in the deep interior, 
the initial dynamical response would quickly stagnate, leaving then
radiative diffusion as the fall-back transport.
But since massive stars are already close to the Eddington limit,
the associated excess luminosity should push it over this limit, 
leading then to the strong, {\em radiatively driven} mass loss 
described above.

Because this time scale is still much longer than any dynamical time
in the system, the
essential processes can be modeled in terms of a quasi-steady
continuum-driven wind during this super-Eddington epoch, 
as described above.

Perhaps the least understood aspect of LBVs is the mechanism giving
rise to the observed eruptions.  In supernovae explosions, the energy
source is obviously the core-collapse to a neutron star or black hole.
But in LBV eruptions, the post-eruption survival of an intact star,
and the indication at least some LBVs can undergo multiple giant
eruptions, both show that the energy source cannot be a one-time
singular event like core collapse. Some other mechanism must provide the energy, 
but the exact nature of this is still unknown.
So it is still unclear why LBVs erupt, or what sets the eruption time, amplitude, and repetition rate.
In particular, there is currently no model that predicts these quantities.


\section{Concluding Summary}

An overall theme of this chapter is that, because of their very high luminosity, radiative forces play an important, dynamical role in the stability of the envelopes and winds of VMS.
A key issue is the nature of the opacity that links the radiation to gas, and in particular the distinction between line vs. continuum processes.
Line opacity can in principle be much stronger, but in the stellar envelope the saturation of the radiative flux within the line means that flux-weighted line-force depends on an inverse or harmonic mean (a.k.a. Rosseland mean).
This  only becomes moderately strong (factor ten above electron scattering) in regions of strong line overlap, most particularly the so-called Iron bump near 150,000~K. This iron bump can cause a strong, even runaway inflation of the stellar envelope, leading to an Iron-Bump Eddington Limit that might be associated with S-Doradus type LBVs.

Near surface layers, the desaturation of the lines leads to a much stronger line-force that drives a strong stellar wind, with a well defined  mass loss rate regulated by the level of line saturation at the sonic point base.
Away from the wind base, there develops a strong "line-deshadowing instability" that induces an extensive  clumping and associated porosity in the outer wind.

For stars that exceed the classical Eddington limit, much stronger mass loss can be driven by the continuum opacity, even approaching the ``photon tiring'' limit, in which the full stellar energy flux is expended to lift and accelerate the mass outflow.
A key issue here is regulation of the continuum driving by the porosity that develops from instability and flow stagnation of the underlying stellar envelope.
For a simple power-law model of the porous structure, the derived mass loss rates seem capable of explaining the giant eruption LBVs, including the 1840's eruption seen in eta Carinae.
Two key remaining issues are the cause of the super-Eddington luminosity, and whether the response might be better modeled as an explosion vs. a quasi-steady mass loss eruption.
  
\begin{acknowledgement}
This work was supported in part by NASA ATP grant NNX11AC40G, NASA Chandra grant TM3-14001A, and NSF grant  1312898 to the University of Delaware. I thank M. Giannotti for sharing his Mathematica notebook for the OPAL opacity tables, and N. Shaviv for many helpful discussions and for providing figure 12. I also acknowledge numerous discussions with G. Graefener, N. Smith, J. Sundqvisit,  J. Vink and A.J. van Marle.

\end{acknowledgement}

\bibliographystyle{mn2e}

\bibliography{OwockiS}

\end{document}